\documentclass[preprint,onecolumn,nofootinbib]{revtex4}
\usepackage[colorlinks=true,linkcolor=blue,urlcolor=blue,filecolor=black,citecolor=red,pdfstartview=FitV,pdftitle={},pdfsubject={},pdfkeywords={},pdfpagemode=None,bookmarksopen=true]{hyperref}
\usepackage{graphicx}%Include figure files
\usepackage{amsmath}
\usepackage{amsfonts}
\usepackage{amssymb,ulem}
\usepackage{color,xcolor}%
\usepackage{CJK}
\usepackage{subfigure}
\usepackage{amsthm,amsmath,amssymb}
\usepackage{mathrsfs}

\usepackage{dcolumn}% Align table columns on decimal point
\usepackage{float}
\usepackage{multirow}% apply to table
\begin{document}
\title{Quasinormal modes and Hawking radiation of a charged Weyl black hole}
\author{Guoyang Fu$^{1}$}
\thanks{FuguoyangEDU@163.com}
\author{Dan Zhang $^{2}$}
\thanks{danzhanglnk@163.com}
\author{Peng Liu $^{3}$}
\thanks{phylp@email.jnu.edu.cn}
\author{Xiao-Mei Kuang$^{1,4}$}
\thanks{xmeikuang@yzu.edu.cn}
\author{Qiyuan Pan$^{2}$}
\thanks{panqiyuan@hunnu.edu.cn}
\author{Jian-Pin Wu$^{1,4}$}
\thanks{jianpinwu@yzu.edu.cn}
\affiliation{$^1$ Center for Gravitation and Cosmology, College of Physical Science and Technology, Yangzhou University, Yangzhou 225009, China}
\affiliation{ $^2$ Key Laboratory of Low Dimensional Quantum Structures and Quantum Control of Ministry of Education, Synergetic Innovation Center for Quantum Effects and Applications, and Department of Physics, Hunan Normal University, Changsha, Hunan 410081, China}
\affiliation{$^3$ Department of Physics and Siyuan Laboratory, Jinan University, Guangzhou 510632, China}
\affiliation{$^4$ Shanghai Frontier Science Center for Gravitational Wave Detection, Shanghai Jiao Tong University, Shanghai 200240, China}
	
\begin{abstract}
\baselineskip=0.7 cm

We investigate the scalar field system over a charged Weyl black hole, depicted by a parameter $\lambda$.
It is found that the imaginary part of the quasinormal mode spectra is always negative and the perturbation does not increase with the time, indicating that the system is stable under scalar field perturbation. Furthermore, the quasinormal mode spectra and Hawking radiation exhibit a qualitatively similar characteristic in that they both rise with rising $\lambda$ and approach a constant when $\lambda$ is large enough. Especially, we would like to emphasize that an exponential decay obviously emerges in the phase of the ringing tail as $\lambda$ increases. It indicates that the characteristic parameter $\lambda$ has an obvious imprint on the ringing tail, which is expected to be detected by future observations.

\end{abstract}
	
\maketitle
\tableofcontents
	
\section{Introduction}

Recent observations of gravitational waves (GWs) from the coalescence of binary systems \cite{LIGOScientific:2016aoc,LIGOScientific:2016lio,LIGOScientific:2016sjg} and shadows of supermassive black holes (M87$^*$ and SgrA$^*$) by the Event Horizon Telescope  \cite{EventHorizonTelescope:2019dse,EventHorizonTelescope:2019ths,EventHorizonTelescope:2022xnr,EventHorizonTelescope:2022xqj} confirm the existence of black hole and thus test the robustness of general relativity (GR). Nonetheless, there are still numerous open fundamental questions, including quantum gravity, dark energy and dark matter problems, and so on.
These unresolved problems have spurred a renewed interest in the gravity theories beyond GR, especially, their effects deviating from GR could have possible prints in the detected GWs and black hole shadows.

An interesting modified gravity theory is the Weyl gravity.
It is a fourth-order gravity theory, originally proposed by H. Weyl \cite{Weyl:1918pdp}. Since the Weyl gravity is power-counting renormalizable \cite{PhysRevD.16.953,Faria:2015vea}, it is a suitable candidate to construct quantum gravity theory \cite{BERGSHOEFF1981173,deWit:1980lyi}. This theory is also a possible UV completion of GR. It is worth pointing out that there is an equivalence between GR and Weyl gravity with the Neumann boundary conditions \cite{Maldacena:2011mk,Anastasiou:2016jix}.

Specifically, the action of the Weyl gravity is
\cite{Weyl:1918pdp,Konoplya:2020fwg}
%%%%%%%ppp
\begin{eqnarray}\label{Weyl_action}
	S= {-\kappa}\int d^{4}x \sqrt{-g} C_{\mu \nu \rho \sigma}C^{\mu \nu \rho \sigma}\,,
\end{eqnarray}
%%%%%%ppp
where $\kappa$ is the coupling constant. The Weyl tensor
%%%%%%ppp
\begin{eqnarray}
	C_{\mu \nu \rho \sigma}=R_{\mu \nu \rho \sigma}+\frac{R}{6}(g_{\mu \rho}g_{\nu \sigma}-g_{\mu \sigma}g_{\nu\rho})-\frac{1}{2}(g_{\mu \rho}R_{\nu\sigma}-g_{\mu\sigma}R_{\nu\rho}-g_{\nu\rho}R_{\mu\sigma}+g_{\nu\sigma}R_{\mu\rho})\,,
	\label{Weyl-tensor}
\end{eqnarray}
is invariant under the local conformal transformation $\tilde{g}_{\mu\nu}=\Omega^2 g_{\mu\nu}$, where $\Omega$ is a function of the local spacetime point. Such transformation preserves the angles but not the distances.
A static and spherically symmetric vacuum black hole solution from Weyl gravity is worked out in \cite{Mannheim:1988dj}. Of particular interest is that it can address both the dark energy related phenomena \cite{Mannheim:2005bfa,Robert:2013Entrp} and the dark matter scenario \cite{Mannheim:1988dj}. Further, the general Reissner-Nordstr\"om (RN), Kerr and Kerr-Newman solutions are also obtained in \cite{Mannheim:1990ya}.

More recently, an alternative  black hole solution from Weyl gravity has been constructed with the use of the background field method and linear approximation \cite{Tanhayi:2011dh}. With this inspiration, a charged Weyl black hole has also been proposed in \cite{Payandeh:2012mj}. This charged solution can be reformed into a RN-like metric, but the sign in the \textquotedblleft charge" term is minus instead. Thus, here we have great interest in disclosing the characterized features of such a RN-like black hole in Weyl gravity.

It is well known that one powerful way to extract the black hole characterization is to perturb it and then see how it responds. Theoretically, to perturb a black hole spacetime, one can either introduce a probe field into the black hole spacetime or perturb the black hole metric itself. The former is simplified to the field propagation in the black hole background when a field does not backreact on the background. The physics after perturbing a black hole is complex, but we know that it will result in the radiation of GWs, and before the system relaxes to be equilibrium, there exists the black hole merger phase in which  the excitation of other matter fields can occur. This stage is known as the ringdown phase, and in this phase the black hole emits the GWs with the characteristic discrete frequencies, dubbed the quasinormal mode (QNM) frequencies that encode the decaying scales and dampened oscillating frequencies \cite{Berti:2009kk}. More importantly, the QNM spectra of the matter fields also depend on the background spacetime. It is expected that any deviation from GR has an imprint on the QNM spectra, thus serving as a specific probe of modified gravity \cite{Berti:2005ys,Berti:2018vdi}. Given all that, as the first step toward understanding the properties of this charged Weyl black hole, we consider a probe massless scalar field over this background and study the properties of its QNM spectra.

We are also interested in the Hawking radiation as a quantum effect, which could partly describe the near horizon nature of a black hole \cite{Hawking:1975}. It is well known that the Hawking radiation is not an ideal black body since the particles
created in the vicinity of event horizon without enough energy cannot penetrate
the potential barrier. So only part of the particles can be observed at infinity and the radiation behaves as a gray body. This process makes the radiation be a scattering problem. Thus, we can just solve the wave equation outside the black hole and calculate the scattering coefficient which could further give us the gray-body factor as well as the energy radiation rate.
Lots of works on the Hawking emission spectra in GR, modified gravity theory, and in the analogue gravity theory, which is aimed at testing the Hawking radiation in the
laboratory, have been widely explored to further understand the features of the black holes (see for example \cite{PhysRevD.13.198,Harris:2003eg,Zhang:2020qam,Konoplya:2019hml,Konoplya:2019ppy,Konoplya:2020jgt,
Konoplya:2020cbv,Konoplya:2021ube,Guo:2020blq,Ling:2021vgk,Syu:2022cws}).

Our paper is organized as what follows. In Sec. \ref{sec-bb}, we briefly review the charged Weyl black hole solution and analyze the instability of this black hole under the scalar field perturbation. Sections \ref{sec-qnms} and \ref{sec-evo} are, respectively, dedicated to the properties of the QNM spectra and the dynamical evolution of the scalar field. Furthermore, Sec. \ref{sec-Hawking} focuses on the gray-body factor and energy emission rate. In Sec. \ref{sec-conclusion}, we present the conclusions and discussions. In addition, we also give a brief introduction on the Wentzel-Kramers-Brillouin(WKB) method in Appendix \ref{WKB method} and discuss how to filter out the spurious modes when we use the pseudospectral method to find the QNMs in Appendix \ref{filter}.

\section{Massless scalar field over charged Weyl black hole}\label{sec-bb}

The RN-like solution of Weyl gravity in the presence of a charged source is given by \cite{Payandeh:2012mj}
%%%%
\begin{eqnarray}
	\label{metric}
	ds^2=-B(r)dt^2+\frac{dr^2}{B(r)}+r^{2}(d\theta^{2}+{\sin^{2}{\theta}}d\phi^{2})\,,
\end{eqnarray}
with
\begin{eqnarray}
	B(r)=1-\frac{r^2}{\lambda^2}-\frac{Q^2}{4r^2}\,,
	\label{metric-Br}
\end{eqnarray}
where $\lambda$ and $Q$ are the black hole parameters. The second term $r^2/\lambda^2$ in the lapse function $B(r)$ is related to the dark energy scenario.

When $\lambda <Q$, a naked singularity is encountered. If $\lambda >Q$, then the spacetime admits two horizons: the event horizon $r_h$ and the cosmological horizon $r_c$, located at \cite{Fathi:2020sey}
%%%%%%ppp
\begin{eqnarray}
	r_h=\lambda \sin\left(\frac{1}{2} \arcsin\left(\frac{Q}{\lambda}\right)\right)\,, \\
	r_c=\lambda \cos\left(\frac{1}{2} \arcsin\left(\frac{Q}{\lambda}\right)\right)\,.
\end{eqnarray}
%%%%%%ppp
Then, we have the Hawking temperature as
\begin{eqnarray}
	T_H=\frac{1}{\sqrt{2}\pi Q \lambda^{3/2}}\Big(\lambda\sqrt{\lambda+\sqrt{\lambda^2-Q^2}}-\frac{Q^2}{\sqrt{\lambda+\sqrt{\lambda^2-Q^2}}}\Big)\,.
\end{eqnarray}
It is easy to find that when $\lambda=Q$, one has an extremal black hole, possessing a unique horizon at $r_{ex}=r_h=r_c=\lambda /\sqrt{2}$.
%%%%%%%%%%
\begin{figure}[H]
	\center{
		\includegraphics[scale=0.78]{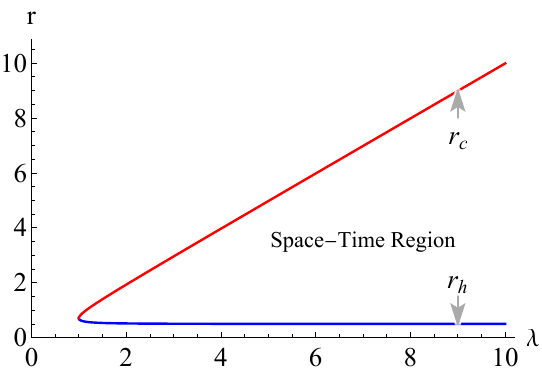}\hspace{0.6cm}
		\includegraphics[scale=0.64]{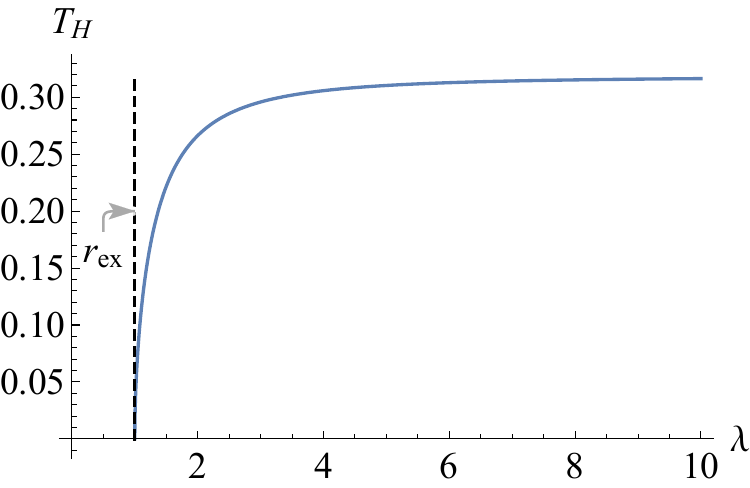}\
		\caption{Left plot: the horizon structure of the charged Weyl black hole for different $\lambda$. Right plot: {the Hawking temperature as the function of $\lambda$.}}
			\label{horizon-tem-vslambda}
		}	
\end{figure}

Because the system is invariant under the following rescaling: $r\to r Q$, $\lambda \to \lambda Q$, and $t \to t Q$, through this paper, we will set $Q=1$ and only leave $\lambda$ free without loss of generality.
Left plot in Fig.\ref{horizon-tem-vslambda} shows the horizon structure of this black hole for different $\lambda$. We clearly see that when $\lambda=Q=1$, the cosmological horizon coincides with the event horizon. With $\lambda$ growing larger, the spacetime region also becomes larger. We also show the Hawking temperature as the function of $\lambda$ in the right plot in Fig.\ref{horizon-tem-vslambda}. We find that
the Hawking temperature increases with $\lambda$, and then approaches a constant as $\lambda\to\infty$, i.e., $T_H \big{|}_{\lambda \to \infty }=1/\pi$.
	
Lots of works based on this black hole background have been widely explored, including the motion of massless particle, neutral massive particle, and electrically charged particle \cite{Fathi:2019jgd,Fathi:2020sey,Fathi:2020otm,Fathi:2020sfw}.
Here, we shall study the QNM spectra, the dynamical evolution of a massless scalar field over this black hole, and also its Hawking radiation, which could help to further understand the properties of the black hole.
	
A probe massless scalar field $\psi$ over the charged Weyl black hole can be described by the Klein-Gordon (KG) equation:
	%%%%%%ppp
\begin{eqnarray} \label{KG_equation}
	\frac{1}{\sqrt{-g}} \partial_\mu(\sqrt{-g}g^{\mu\nu}\partial_\nu\psi)=0\,.
\end{eqnarray}
	%%%%%%ppp
After making a separation of variables by a spherical harmonic $\psi=\Psi(t,r)Y_l(\theta, \phi)/r$, we can recast the KG equation into the Schr$\ddot{o}$dinger-like form
	%%%%%%ppp
\begin{eqnarray}\label{TD_equation}
	-\frac{\partial^2 \Psi}{\partial t^2}+\frac{\partial^2 \Psi}{\partial r_*^2}-V(r)\Psi=0\,,
\end{eqnarray}
	%%%%%%ppp
where $r_*$ is the tortoise coordinate defined as $dr_*=dr/B(r)$. The tortoise coordinate's analytic form can be also explicitly written as
\begin{eqnarray}\label{rstart}
r_*&=&-\frac{r_c(r_c^2+r_h^2)}{2(r_c^2-r_h^2)}\log\left(1-\frac{r}{r_c}\right)+ \frac{r_h(r_c^2+r_h^2)}{2(r_c^2-r_h^2)}\log\left(\frac{r}{r_h}-1\right)
	\nonumber
\\
&&
-
	\frac{r_h(r_c^2+r_h^2)}{2(r_c^2-r_h^2)}\log\left(1+\frac{r}{r_h}\right)
	+
	\frac{r_c(r_c^2+r_h^2)}{2(r_c^2-r_h^2)}\log\left(1+\frac{r}{r_c}\right)\,.
\end{eqnarray}
$V(r)$ is the effective potential
	%%%%%%ppp
\begin{eqnarray}\label{effective_V}
	V(r)=B(r)\left(\frac{l(l+1)}{r^2}+\frac{B'(r)}{r}\right)\,.
\end{eqnarray}
	%%%%%%ppp
Here $l$ is the angular quantum number. This effective potential  obviously depends on the black hole background as well as the angular quantum number.

\begin{figure}[H]
	\center{
		\includegraphics[scale=0.8]{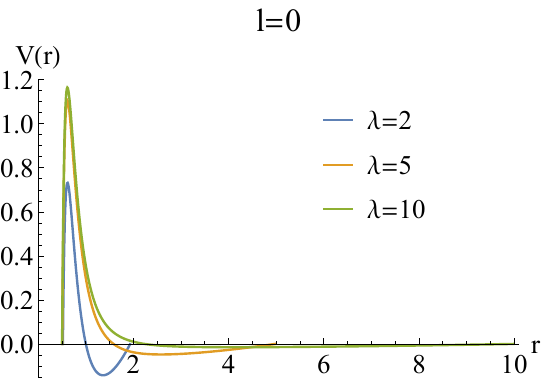}\hspace{0.5cm}
		\includegraphics[scale=0.8]{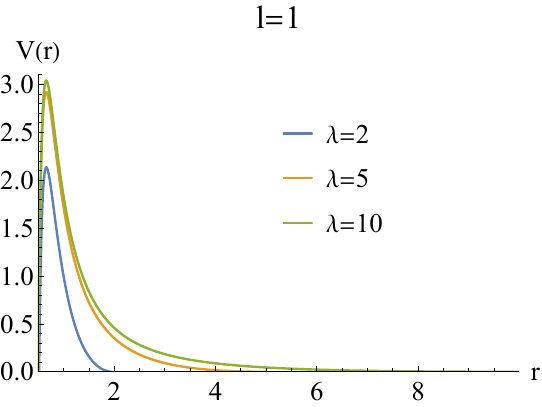}\
		\caption{The effective potential $V(r)$ for different $\lambda$ with fixed $l$.}
		\label{instability_l0}
	}
\end{figure}

Left plot in Fig.\ref{instability_l0} shows the effective potential for $l=0$ with different $\lambda$. A negative gap can be observed in the effective potential. The negative gap is a probable indicator of instability, although not always. It advises a rigorous examination of the QNMs and the time domain profile to corroborate the stability characteristic. This will be illustrated in the sections that follow.
While for $l>0$, the effective potentials are always positive (the right plot in Fig.\ref{instability_l0}), indicating that the system is stable under the perturbation of the scalar field.
%%%%%%%%%%%
%%%%%%%%%%%

\begin{figure}[H]
	\center{
		\includegraphics[scale=0.8]{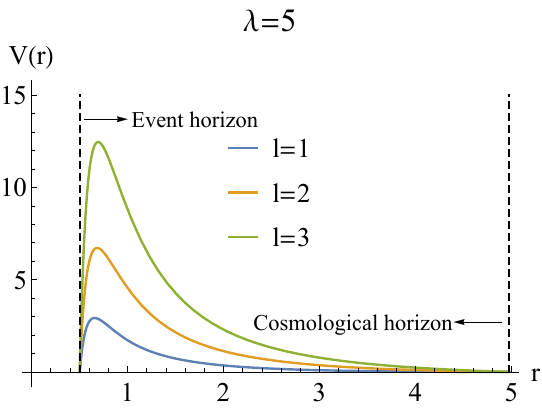}\hspace{0.1cm}
		\includegraphics[scale=0.8]{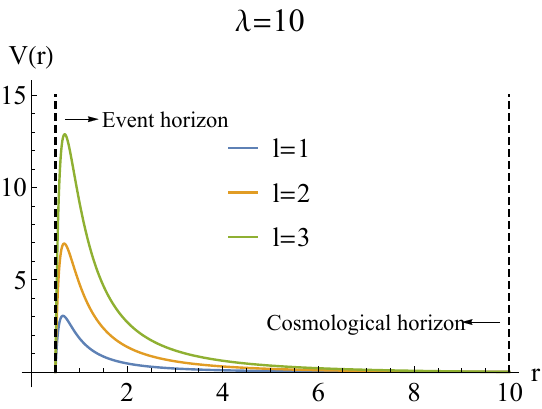}\hspace{0.1cm}
		\caption{The effective potential $V(r)$ for different $l$ with fixed $\lambda$.}
		\label{effective_potential}
	}
\end{figure}

Further, from Fig.\ref{effective_potential}, we observe that for fixed black hole parameter $\lambda$, the height of the potential barrier grows with the angular quantum number. This is the universal property of the black hole potential barrier. For fixed $l$, we see that the height of the potential barrier grows with increasing $\lambda$ (see Fig.\ref{instability_l0}). However, once $\lambda$ grows large enough, the increase in height becomes moderate.
There is no doubt that the shape of the effective potential shall make a significant impact on the QNMs, dynamical evolution and Hawking radiation, which shall be illustrated in what follows.

%%%%%%%%%%
%%%%%%%%%%%

\section{Quasinormal modes}\label{sec-qnms}
	
QNMs are an intrinsic characteristic of the background spacetime, and therefore its spectra encode the key information about black holes \cite{Chandrasekhar,Nollert:1999ji,Berti:2009kk,Konoplya:2011qq,Kokkotas:1999bd}. The nature of determining the QNMs is to solve the eigenvalue problem. There are several methods developed to determine the QNMs, among which the pseudospectral method is one of the powerful numerical tools. In this section, we shall implement pseudospectral method to calculate the QNM spectra. For the pseudospectral method, we can refer to \cite{Boyd:Chebyshev} and also see \cite{Jansen:2017oag,Wu:2018vlj,Fu:2018yqx,Xiong:2021cth,Liu:2021fzr,Liu:2021zmi,Jaramillo:2020tuu,Jaramillo:2021tmt,Destounis:2021lum} for the application in the calculation of QNMs in black hole physics. We justify our findings by further cross-checking the results with the WKB method, which is a widely used and well-understood method, and analyze the error between the WKB and pseudospectral methods. A brief introduction on the WKB method shall be presented in the Appendix \ref{WKB method}.
	
The key point of the pseudospectral method is to discretize the differential equations and then solve the resulting generalized eigenvalue equations. Specially, we replace the continuous variables by a discrete set of collocation points called the grid points and expand the functions by some particular basis functions called cardinal functions.
Usually, we use the Chebyshev grids and Lagrange cardinal functions
\begin{eqnarray}
		\label{Cg-Lcf}
		x_i=cos(\frac{i}{N}\pi)\,, \ \ C_j(x)=\prod_{j=0,j\neq i}^N \frac{x-x_j}{x_i-x_j}\,, \ i=0\,, ...\,, N\,.
\end{eqnarray}
	
Now we are ready to determine the QNM spectra. Let us first expand $\Psi$ as $\Psi=e^{-i\omega t}\Phi$ such that we work in the frequency domain. Then, the KG equation takes the form
	%%%%%%ppp
\begin{eqnarray} \label{QNMs_equationv1}
		\frac{\partial^2 \Phi}{\partial r_*^2}+(\omega^2-V(r))\Phi=0\,.
\end{eqnarray}
	%%%%%%ppp	
At the boundaries, one has
\begin{eqnarray}
	\Phi\sim e^{\pm r_*}\,,\,\,\,\,\, r_*\rightarrow\pm\infty\,,
\end{eqnarray}
which correspond to a vanishing wave function. The aforementioned boundary conditions indicate that the waves are purely outgoing at infinity and purely ingoing on the event horizon, implying that no waves from the horizon or infinity are permitted. These boundary conditions reflect a black hole's response to a transient perturbation, after the source has ceased to act \cite{Konoplya:2011qq,Berti:2009kk,Kokkotas:1999bd}.

To calculate the QNM spectra, it is convenient to work in the Eddingtton-Finkelstein coordinate, where Eq.\eqref{QNMs_equationv1} is linear in the frequency $\omega$. And then, one obtains the generalized eigenvalue equation as
	%%%%%%ppp
	\begin{eqnarray}\label{generalized_eq}
		(M_0+\omega M_1)\Phi=0\,,
	\end{eqnarray}
	%%%%%%ppp	
where $M_i$ ($i=0,1$) are the linear combination of the derivative matrices. The above equation can be solved directly by the Eigenvalue function in the $Mathematica$\footnote{In order to solve Eq.\eqref{generalized_eq}, we need to impose the proper boundary conditions near the cosmological horizon. For more details, please refer to Ref.\cite{Jansen:2017oag}.}. It is found that there is an infinite but discrete set of eigenfrequencies $\omega_{ln}$, where $l$ is the angular momentum number and $n$ is the overtone number characterizing the number of nodes of the radial solution.

%%%%%%%%%%
\begin{figure}[H]
	\center{
		\includegraphics[scale=0.35]{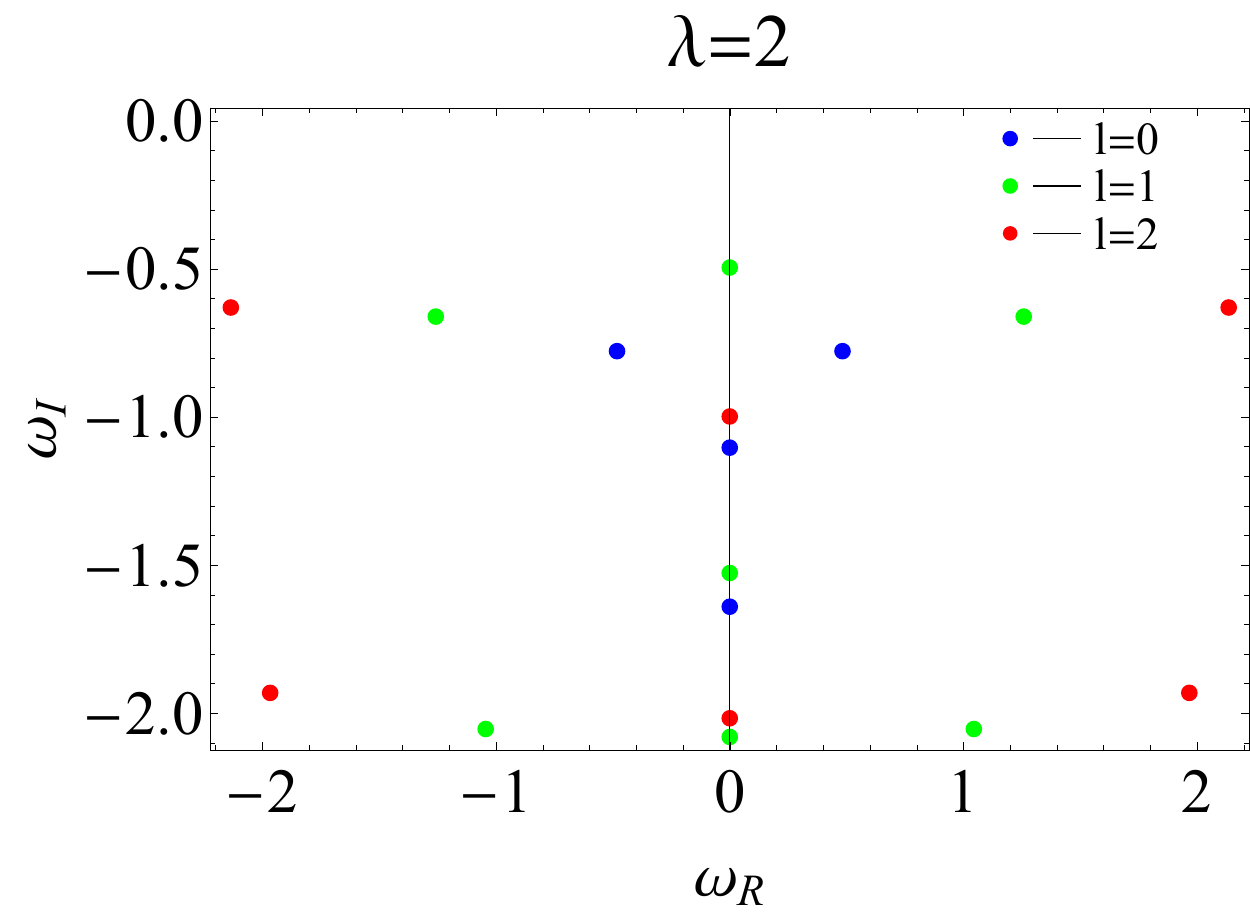}\hspace{0.1cm}
		\includegraphics[scale=0.35]{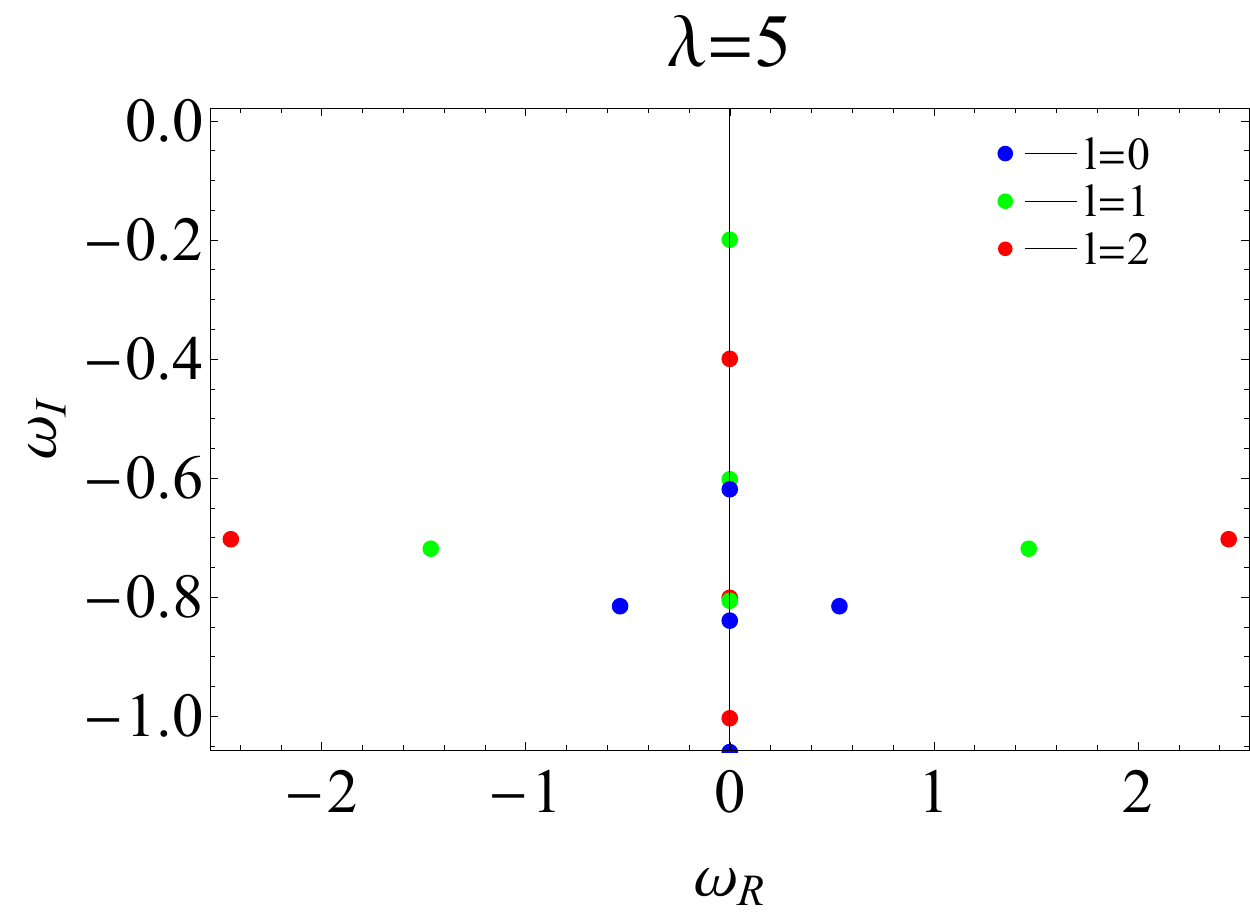}\\
		\caption{The pole structures of the QNM spectra for different $\lambda$ and $l$.}
		\label{Fig_QNMs}
	}
\end{figure}
%%%%%%%%%%%

The QNM spectra are shown in Fig.\ref{Fig_QNMs}\footnote{We have filtered out the spurious modes and only present the genuine modes in this figure and the figures and tables following in the main body. Please see Appendix \ref{filter} for more information on how to discriminate and filter the spurious modes.}.
The imaginary part of the QNM spectra is always negative, indicating that the system is stable under scalar field perturbation. 
Particularly noteworthy is the discovery of two families of modes in the QNM spectra: photon sphere (PS) mode and de Sitter (dS) mode. PS mode may be traced back to the photon sphere at the large $l$ and is well described by conventional WKB-type methods \cite{Cardoso:2017soq}. The dS mode is a pure imaginary mode whose existence and timescale is intrinsically linked to the de Sitter horizon, according to \cite{Cardoso:2017soq}. Conventional WKB-type approaches fail to locate the dS mode \cite{Konoplya:2022gjp}. The pseudospectral approach outlined above is an effective tool for locating such modes.

\begin{figure}[H]
	\center{
		\includegraphics[scale=0.75]{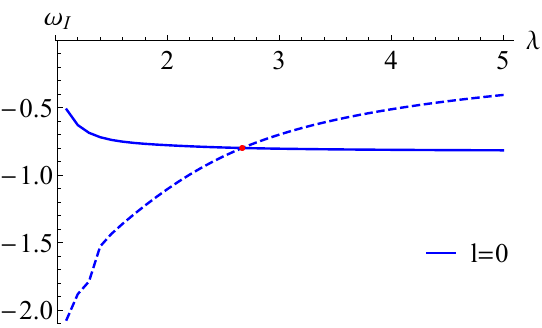}\hspace{0.8cm}
		\includegraphics[scale=0.75]{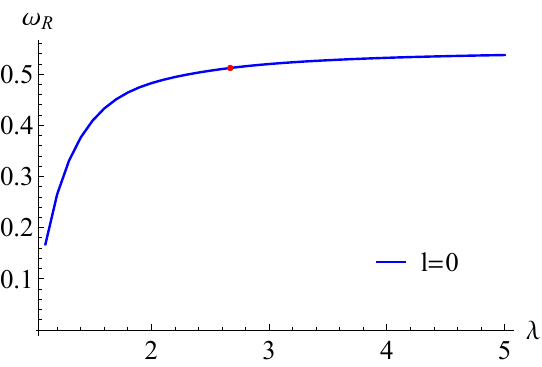}\\
		\includegraphics[scale=0.6]{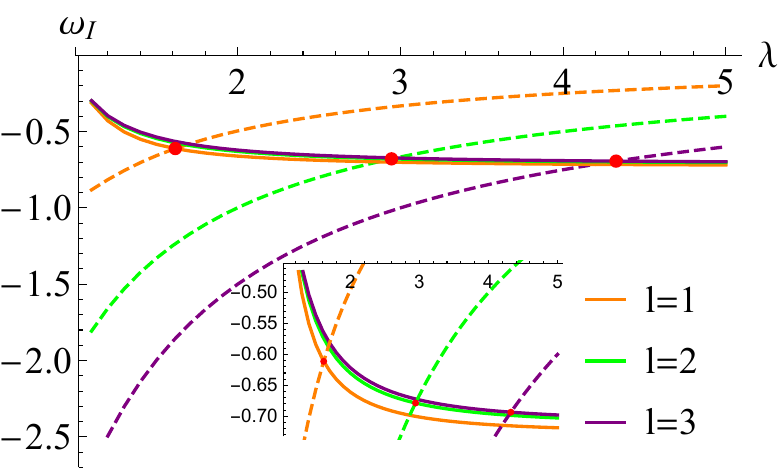}\hspace{0.8cm}
		\includegraphics[scale=0.75]{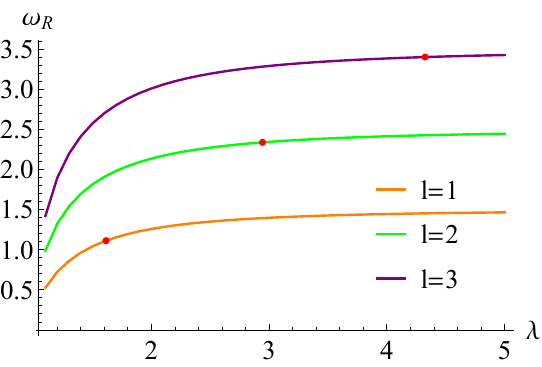}\\
		\caption{The photon sphere mode (solid line) and dS mode (dashed line) as a function of $\lambda$ for several $l$.}
		\label{Fig_QNMsv1}
	}
\end{figure}

In Fig.\ref{Fig_QNMsv1}, we describe the PS mode (solid line) and dS mode (dashed line) as a function of $\lambda$ for several $l$. It is evident that the PS mode is dominant for small $\lambda$. The imaginary part of the PS mode diminishes as $\lambda$ grows, but the dS mode increases. As a result, there is a critical value $\lambda_c$ (red dots in Fig.\ref{Fig_QNMsv1}) beyond which the dS mode takes precedence over the PS mode. The similar behavior of the QNM frequency is also found in the Schwarzschild-dS black hole \cite{Zhidenko:2003wq,Jansen:2017oag}. It attributes to the fact that the $\lambda$ plays a role as the inverse proportion of cosmological constant $\Lambda$. We also see that when $\lambda$ is large enough, both the real and imaginary parts of the PS mode or the dS mode approach a constant. This discovery is compatible with the fact that the potential barrier, particularly its height, is almost the same for large $\lambda$ (only very small difference).

%%%%%%%%%%
%%%%%%%%%%%

Finally, we would like to compare the results of the QNM spectra obtained by the pseudospectral technique and the sixth order WKB method\footnote{We will demonstrate the choice of order of WKB method in Appendix \ref{WKB method}.} and discuss the errors between the two approaches. Because the WKB approach fails to find the dS mode, we only discuss the PS mode results here. The results are displayed in Table \ref{tablev1}. This table also includes the errors evaluated by $\delta \omega_{error}=|\omega_{PS}-\omega_{WKB_6}|/2$ between the two approaches. For fixed $n$, it is evident that the errors reduce as $l$ grows. It is consistent with the claim that the WKB formula usually provides better accuracy for larger $l$ ($l> n$) than smaller $l$\footnote{Readers can refer to Refs.\cite{Konoplya:2003ii,Konoplya:2019hlu}, as well as the discussions in Appendix \ref{WKB method}.}.

	%%%%%%%%%%%%%%%%%%%%%%%%%%%%%%%%%%%%%
\begin{table}[H]
	\centering
	\begin{tabular}{|c|cc|c|c|c|c|c|}
		\hline
		$n$          &  $l$  &  \multicolumn{2}{|c|}{PS}                    &  \multicolumn{2}{|c|}{WKB$_6$}              & $\delta \omega_{error}$  \\
		\hline
		\multirow{3}{*}{0} &   1   &  \multicolumn{2}{|c|}{1.4650065-0.7187528i}  &  \multicolumn{2}{|c|}{1.4647270-0.718787i}  &       0.000140805        \\
		\cline{2-7}
		&   2   &  \multicolumn{2}{|c|}{2.4447382-0.7028046i}  &  \multicolumn{2}{|c|}{2.4447097-0.702766i}  &       0.0000238985       \\
		\cline{2-7}
		&   3   &  \multicolumn{2}{|c|}{3.4254973-0.6980106i}  &  \multicolumn{2}{|c|}{3.4254982-0.6980142i}  &          $1.82346*10^{-6}$ \\       \hline
		\multirow{3}{*}{1} &   1   &  \multicolumn{2}{|c|}{1.1221779-2.3591926i}  &  \multicolumn{2}{|c|}{1.1198092-2.3710473i} &       0.00604452        \\
		\cline{2-7}
		&   2   &  \multicolumn{2}{|c|}{2.1875304-2.1898062i}  &  \multicolumn{2}{|c|}{2.1853061-2.1928363i} &       0.00187944       \\
		\cline{2-7}
		&   3   &  \multicolumn{2}{|c|}{3.2324728-2.1358198i}  &  \multicolumn{2}{|c|}{3.2322875-2.1359638i} &          0.00011733    \\
		\hline 	
	\end{tabular}
	\label{tablev1}
	\caption{QNM spectra with $\lambda=5$ for various angular number $l$ and overtone number $n$ obtained by pseudospectral method and sixth order WKB (WKB$_6$) method.}
\end{table}
%%%%%%%%%%%%%%%%%%%%%%%%%%%%%%%%%%%%%%%

\section{Dynamical evolution}\label{sec-evo}
	
\begin{figure}[H]
		\center{
			\includegraphics[scale=0.6]{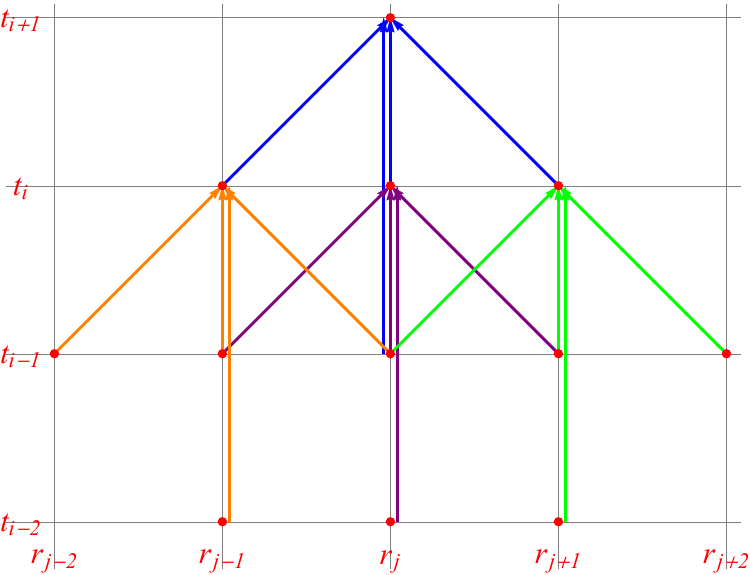}\  \hspace{0.4cm}
			\includegraphics[scale=0.6]{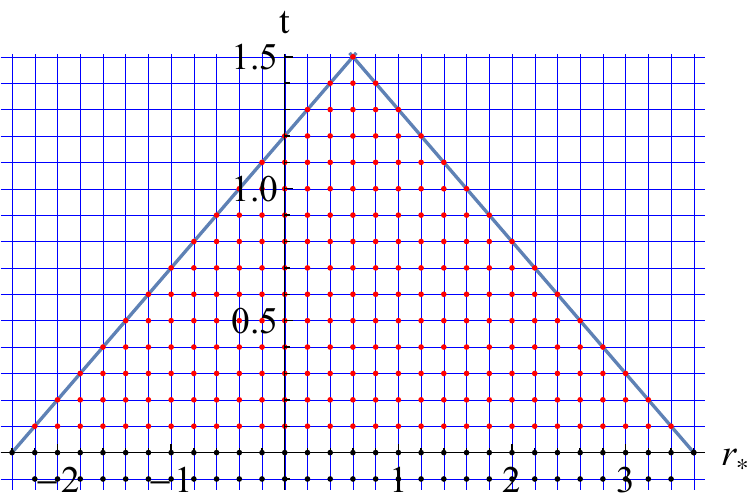}\\
			\caption{The cartoon diagram of FDM. Left plot describes the discretization scheme of the coordinates $(t,r_*)$. Right plot describes the iterative process of the FDM. The black points are the grids to which the initial conditions are assigned. Then, the red points can be evaluated by the iterative process [Eq.\eqref{it-process}].}
			\label{FDM}
		}
	\end{figure}

In this section, we shall explore the dynamical evolution of the massless scalar field for given initial perturbation. We are specially interested in the behaviors of the scalar field in the ringdown phase.
The finite difference method (FDM) is a suitable approach to implement the dynamical evolution.
Before proceeding, we briefly outline the key point of the FDM. For the details, we can refer to Refs.\cite{Abdalla:2010nq,Zhu:2014sya,Lin:2022owb}.
First, we need to discretize the wave equation \eqref{TD_equation}. The discretization scheme is to define $\Psi(r_*,t)=\Psi(j\triangle r_*, i\triangle t)=\Psi_{j,i}$ and $V(r(r_*))=V(j\triangle r_*)=V_j$ [see the left plot in Fig.\ref{FDM} for the cartoon diagram of the descretization scheme of the coordinates $(t,r_*)$]. Therefore, instead of the differential equation \eqref{TD_equation}, we have the following difference equation:
	%%%%%%ppp
	\begin{eqnarray}
		-\frac{(\Psi_{i+1,j}-2\Psi_{i,j}+\Psi_{i-1,j})}{\triangle t^2}+\frac{(\Psi_{i,j+1}-2\Psi_{i,j}+\Psi_{i,j-1})}{\triangle r_*^2}-V_j \Psi_{i,j}+\mathcal{O}(\triangle t^2)+ \mathcal{O}(\triangle r_*^2)=0 \,. \nonumber \\
	\end{eqnarray}
	%%%%%%ppp
	Given the initial Gaussian distribution $\Psi(r_*,t<0)=0$ and $\Psi(r_*,t=0)=exp[-\frac{(r_*-a)^2}{2b^2}]$ with $a$ and $b$ being the constants, the iterate formula is derived as
	\begin{eqnarray}
		\Psi_{i+1,j}=-\Psi_{i-1,j}+\frac{\triangle t^2}{\triangle r_*^2}(\Psi_{i,j+1}+\Psi_{i,j-1})+(2-2\frac{\triangle t^2}{\triangle r_*^2}-\triangle t^2 V_j)\Psi_{i,j}\,.
		\label{it-process}
	\end{eqnarray}
	The cartoon diagram of the iterative process of FDM is shown in the right plot in Fig.\ref{FDM}.
	The Courant-Friedrichs-Lewy condition for stability requires $\triangle t/ \triangle r_*<1$. Here, we use $\triangle t/ \triangle r_*=0.5$. Actually, the numerical accuracy not only depends on the ratio of $\triangle t/ \triangle r_*$ but also the respective values of $\triangle t$ and $\triangle r_*$. Therefore, we must require $\triangle t$ and $\triangle r_*$ to be small enough to satisfy the precision requirement.
	%%%%%%%%%%
	%%%%%%%%%%%	

\begin{figure}[H]
	\center{
		\includegraphics[scale=0.8]{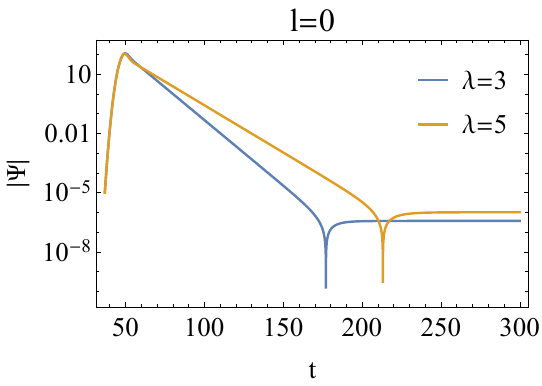}\ \hspace{0.4cm}
		\includegraphics[scale=0.8]{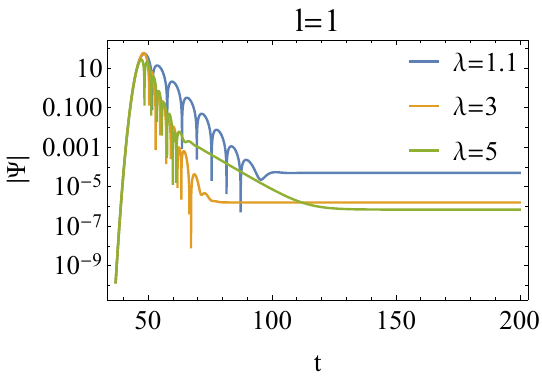}\ \hspace{0.4cm}
		\caption{Semilogarithmic plots of the dynamical evolution of scalar field for varying $\lambda$. }
		\label{Fig_TD}
	}
\end{figure}

We implement the dynamical evolution of the scalar field with varying $\lambda$ for $l=0$ and $l=1$, which are depicted in Fig.\ref{Fig_TD} using the FDM previously stated. It is evident that the perturbation does not increase with the time evolution. This indicates that the system is stable in the presence of a massless scalar field perturbation. 
Then, it is clearly observed that there are two distinct phases for the $l=1$ (see the right plot in Fig.\ref{Fig_TD}): the Schwarzschild-like ringing phase and the de Sitter phase.
The Schwarzschild-like ringing phase exhibits an oscillating tail enveloped by a universal power law decay. While the de Sitter phase is characterized by an exponential tail following the decay law:
	%%%%%%
\begin{eqnarray}
	\left|\Psi\right|\sim\left|\Psi_0\right|+\left|\Psi_1\right|e^{-p_l t}\,,\,\,\,\,\,l=0, 1,\ldots\,,
	\label{decay-behavior}
\end{eqnarray}
where $p_l$ in the above decay law depends on the black hole parameter $\lambda$ as well as the angular number $l$. 
The above decay behavior is similar to that found in the neutral Weyl black hole \cite{Konoplya:2020fwg} but differs from that observed in the usual Schwarzschild-dS black hole \cite{Brady:1996za,Brady:1999wd,Molina:2003dc}, where the constant  $|\Psi_0|$ term occurs only for $l=0$. When $\lambda$ is small (see the right plot in Fig.\ref{Fig_TD} for $\lambda=1.1$), after the Schwarzschild-like ringing phase, the scalar field rapidly evolves into the time independent stage. It suggests that, at this stage, the first term dominates over the second term.
But for large $\lambda$ (see the right plot in Fig.\ref{Fig_TD}), there is an obvious stage following the pure exponential decay as $e^{-p_lt}$ after the Schwarzschild-like ringing phase. After that, the scalar field also evolves into the time independent stage. Therefore, we conclude that there is a larger $p_l$ for small $l$ leading to a more fast decay. As $\lambda$ increases, $p_l$ becomes smaller such that, before entering into the time independent stage, the exponential decay obviously emerges. It indicates that the characteristic parameter $\lambda$ has the imprint on the ringing tail, which is expected to be detected by future observations.
	%%%%%%%%%%
	%%%%%%%%%%%

We would like to point out that this decay law of massless scalar field \eqref{decay-behavior} has been revealed in the neutral Weyl black hole \cite{Konoplya:2020fwg}, comparing to which the effective dark matter ringing phase is absent in our present model. This is reasonable because in this RN-like black hole, the dark matter related term is excluded.
	
\section{Gray-body factor and energy emission rate}\label{sec-Hawking}

The classical radiation dominated by QNMs is studied above. It would be interesting to investigate the quantum radiation, i.e., Hawking radiation of this system further and do comparison between them, which could help to reveal some intrinsic characteristics of this background spacetime and also shed light on the nature of quantum gravity.	

There are several approaches proposed to study the Hawking radiation for
black hole. As we mentioned in the introduction,  we can start from
the wave equation \eqref{QNMs_equationv1} in the frequency domain to
obtain the scattering coefficient from which we can obtain the gray-body factor. Then we
use the gray-body factor to describe the transmission of particles through the potential,
and thus work out the energy radiation rate.
From the above description, in the study we should allow the incoming waves from infinity, such that we evaluate the fraction of particles reflected back from the effective potential barrier to the event horizon. Therefore, contrary to the QNM case, we shall impose the following scattering boundary conditions for Eq.\eqref{QNMs_equationv1}
	\begin{eqnarray}	
		&&
		\Phi=T e^{-i \omega r_*}\,, \ \ \ \ r_* \to - \infty\,,
		\
		\\
		&&
		\Phi=e^{-i \omega r_*}+ R e^{i \omega r_*}\,, \ \  \ \  r_* \to \infty\,,
	\end{eqnarray}
	where $T$ and $R$ are the transmission and reflection coefficients, respectively. They satisfy
	\begin{eqnarray}	
		|T|^2+|R|^2=1\,.
	\end{eqnarray}
	Then, we apply the WKB method to evaluate the reflection coefficient:
	%%%%
	\begin{eqnarray}	
		R=(1+e^{-2i\pi \mathcal{K}})^{-1/2}\,.
		\label{R-coe}
	\end{eqnarray}
	$\mathcal{K}$ in the above expression is determined by
	\begin{eqnarray}	
		\mathcal{K}=i \frac{\omega^2-V_0}{\sqrt{-2 V_2}}-\sum_{k=2}^{k=6} \Lambda_k(\mathcal{K})\,,
	\end{eqnarray}
	where $V_0$ and $V_2$ are the maximal value of the effective potential and its second derivative with respective to $r_*$ at the position of the maximum, respectively. $\Lambda_k(\mathcal{K})$ are the higher WKB correction terms, which only depend on $\mathcal{K}$ and the derivative of the effective potential at the position of its maximum. For the details, please refer to Refs.\cite{PhysRevD.35.3621,1985ApJ291L33S,Konoplya:2003ii} and also Appendix \ref{WKB method}. Here, we evaluate the WKB approach up to the sixth order. Then, one can work out the gray-body factor $|A_l|$ for each angular number $l$
	%%%%
	\begin{eqnarray}	
		|A_l|^2=1-|R|^2=|T|^2 \,.
		\label{grey-body}
	\end{eqnarray}
	
With the gray-body factor at hand, we can study the energy emission rate. We assume that the Hawking temperature of the black hole does not change between the emissions of two consequent particles, which corresponds to the canonical ensemble \cite{Kanti:2004nr}. Then, the energy emission rate has the following form \cite{Hawking:1975vcx}:
	%%%%%
\begin{eqnarray}	
	\frac{dE}{dt}=\sum_l N_l |A_l|^2 \frac{\omega}{exp(\omega /T_H)-1} \frac{d\omega}{2\pi}\,.
\end{eqnarray}
$N_l$ are the multiplicities satisfying $N_l=2l+1$ for the scalar field.

\begin{figure}[H]
	\center{
		\includegraphics[scale=0.72]{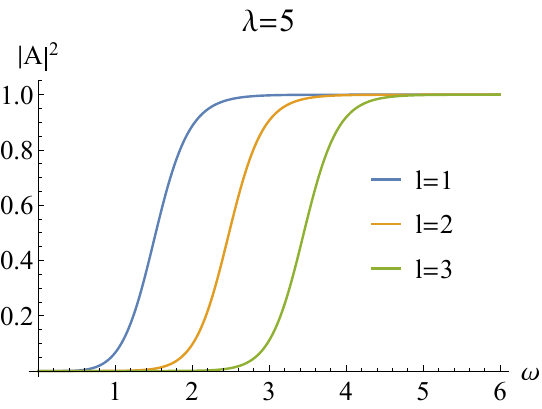}\ \hspace{0.8cm}
		\includegraphics[scale=0.72]{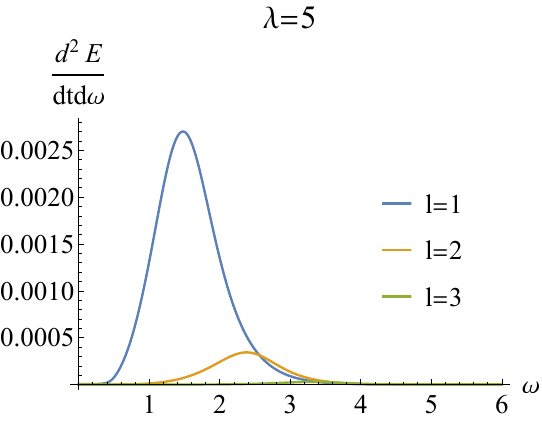}\ \\
		\caption{The gray-body factor (left plots) and the Hawking radiation (right plots) as the function of $\omega$ for fixed $\lambda$ and different $\l$.}
		\label{grey_body_lam}
	}
\end{figure}

\begin{figure}[H]
		\center{
			\includegraphics[scale=0.72]{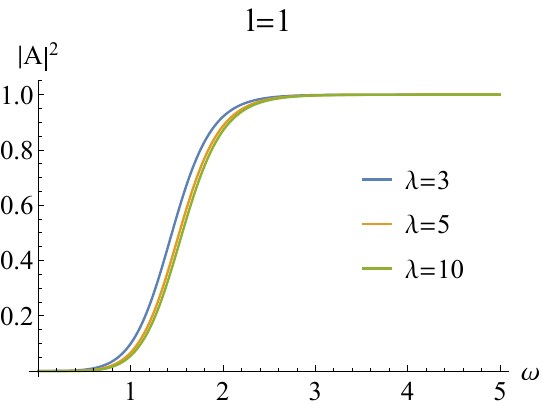}\ \hspace{0.8cm}
			\includegraphics[scale=0.72]{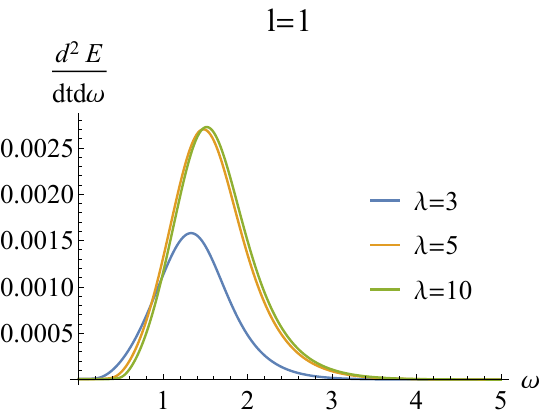}\ \\
			\caption{The gray-body factor (left plots) and the Hawking radiation (right plots) as the function of $\omega$ for fixed $\l$ and different $\lambda$.}
			\label{grey_body_l}
		}
\end{figure}

The numerical results of the gray-body factor and the Hawking radiation are shown in Figs.\ref{grey_body_lam} and \ref{grey_body_l}. We summarize the main properties as follows.
	%%%%%%%%%%
	%%%%%%%%%%%	
	%%%%%%%%%%
	%%%%%%%%%%%	
\begin{itemize}
	\item As the frequency $\omega$ increases, the gray-body factor grows from almost zero to the unit. It is because when the particles possess larger energy, then the probability of penetrating the potential barrier increases.
	\item In the intermediate frequency region, we can obviously see that, for fixed frequency, the smaller the angular number, the larger the gray-body factor (left plots in Fig.\ref{grey_body_lam}). It is because the effective potential has higher barrier for larger $l$ (see Fig.\ref{effective_potential}), which makes it harder for the particles to penetrate the potential barrier forming the transmission of radiation.
	\item With increasing $\lambda$, the gray-body factor slightly decreases for fixed frequency when the frequency is in the intermediate frequency region (left plots in Fig.\ref{grey_body_l}). This result can be also explained by the effective potential as argued in the above point (Fig.\ref{instability_l0}).
	\item The energy emission rate of Hawking radiation is dominated by the modes with lower $l$. The contribution from the modes with higher $l$ is virtually invisible (see the right plot in Fig.\ref{grey_body_lam}).
	For fixed $l$, the energy emission rate of Hawking radiation grows with increasing $\lambda$. But when $\lambda$ is large, the energy emission rate is almost the same (the right plot in Fig. \ref{grey_body_l}).
\end{itemize}	

Finally, we would like to point out that QNMs and Hawking radiation have a qualitatively similar behavior in that they both rise with rising $\lambda$ and approach a constant when $\lambda$ is large enough. To further reveal the differences between these two types of radiation, more quantitative research is called for.

\section{Conclusion and discussion}\label{sec-conclusion}
	
In this paper, we study the properties of QNMs, dynamical evolution and Hawking radiation of a charged Weyl black hole by a probe massless scalar field. The imaginary part of the QNM spectra is always negative and the perturbation does not increase with the time evolution, indicating that the system is stable under scalar field perturbation. We summarize the main properties as what follows.
\begin{itemize}
	\item QNM spectra are classified into two families: PS mode and dS mode. The PS mode is dominant for small $\lambda$. As $\lambda$ increases, the imaginary part of the PS mode decreases, whereas the dS mode rises. Therefore, there is a critical value $\lambda_c$ over which the dS mode takes precedence over the PS mode. It can be attributed to the fact that the $\lambda$ plays a role as the inverse proportion of cosmological constant.
	\item When $\lambda$ becomes large enough, both the real and imaginary parts of the PS mode or the dS mode approach a constant. This discovery is consistent with the fact that the potential barrier, particularly its height, is almost the same for large $\lambda$.
	\item The dynamical evolution of the scalar field consists of two stages: the Schwarzschild-like ringing phase and the de Sitter phase. The effective dark matter ringing phase observed in the neutral Weyl black hole is absent in this charged Weyl black hole background. It can attribute to the dark matter related term is excluded in our present model. Especially, we would like to emphasize that the characteristic parameter $\lambda$ has the imprint on the ringing tail, which is anticipated to be detected by future observations.
	\item In the low frequency region, the gray-body factor vanishes. As the frequency $\omega$ increases, the gray-body grows, and approaches the unit in the high frequency region. This picture is independent of the black hole parameter and the angular number. It can be explained that, as in the low or high frequency region, the effect from the energy dominates over that from the black hole parameter and the angular number. Correspondingly, the Hawking radiation increases with increasing $\omega$ at first, and then decreases after climbing up a maximum. In addition, the energy emission rate of Hawking radiation grows with increasing $\lambda$, and approaches a constant when $\lambda$ is large enough.
	\item The classical radiation dominated by QNMs and the quantum radiation, i.e., Hawking radiation, have a qualitatively similar characteristic in that they both rise with rising $\lambda$ and approach a constant when $\lambda$ is large enough. More quantitative research is still absent. We intend to address this issue in the future so that we can learn more about the differences between these two types of radiation.
\end{itemize}

This work is the first step towards studying the characteristics of the charged Weyl black hole by perturbing the black hole spacetime. It would be intriguing to extend our research to the probe Maxwell and Dirac fields and further explore the response of Weyl black hole. As previously shown, see \cite{Lagos:2020oek,Aragon:2020teq,Fontana:2020syy} and references therein, the decay timescales of the QNMs of a massive scalar field or Dirac field exhibit an anomalous behavior. Depending on the mass of the scalar field or the Dirac field, they either grow or decay with an increasing angular number. From a viewpoint, it is interesting to investigate how the characteristic parameter $\lambda$ affects this anomalous behavior. There is no doubt that studying the gravitational perturbations is more essential since it reflects the fingerprints of GWs. It is also worth investigating the scalarization of this charged Weyl black hole. Typically, scalarization is triggered by an instability induced by the scalar field perturbation \cite{Doneva:2017bvd,Silva:2017uqg}. This study shows that the system is stable under free scalar field perturbation. In order to implement the scalarization of this charged Weyl black hole, we may need to incorporate the nominimal coupling function between the Weyl term and scalar field following the idea in \cite{Doneva:2017bvd,Silva:2017uqg,Herdeiro:2018wub,Yang:2021yoe}. We will investigate these issues in the near future.
	
\acknowledgments
	
This work is supported by National Key R$\&$D Program of China (Grant No. 2020YFC2201400), Natural Science
Foundation of China under Grants No. 12035005, No. 11905083, No. 11775036, No. 12147209 and No. 12275079, Postgraduate Research \& Practice Innovation Program of Jiangsu Province under Grant No. KYCX20\_2973, Fok Ying Tung Education Foundation under Grant No. 171006, Natural Science Foundation of Jiangsu Province under Grant No.BK20211601, the Science and Technology Planning Project of Guangzhou (202201010655) and Postgraduate Scientific Research Innovation Project of Hunan Province  under Grant No. CX20220509. J.-P.W. is also supported by Top Talent Support Program from Yangzhou University.	

\appendix
\section{WKB METHOD AND ERROR ESTIMATION} \label{WKB method}
	
There are several methods determining the QNMs. The WKB method is a widely used and economic semianalytic method to solve the eigenvalue problem. However, the WKB formula usually gives a best accuracy for $l> n$ \cite{Konoplya:2003ii,Konoplya:2019hlu}. When $l\leq n$, this method does not always give a reliable result \cite{Konoplya:2003ii,Konoplya:2019hlu}. In addition, increasing the WKB order does not always gives a better approximation for the QNM spectra. Sometimes higher-order formula increases the error \cite{Konoplya:2003ii,Konoplya:2019hlu}.
	
For a wavelike equation with a potential barrier, when the two turning points close enough, the potential function can be expanded by the Taylor series at the position of the peak of the potential. The key point of WKB method is to match the exterior WKB solutions across the two turning points. Therefore, the validity of this method relies heavily upon the form of the effective potential.
	
The first order WKB was first proposed by Schutz and Will \cite{1985ApJ291L33S}. Then, Iyer and Will developed the third order WKB method \cite{Iyer:1986np,Guinn:1989bn}. The accuracy of the third WKB for the fundamental mode has reached about $1\%$.
Soon afterwards, the WKB method was extended to the sixth order by Konoplya \cite{Konoplya:2004ip,Konoplya:2003ii} and 13th order by Matyjasek and Opala \cite{Matyjasek:2017psv}.
	
Usually, we have the following general higher order WKB formula \cite{Konoplya:2019hlu}
	%%%%%%ppp
\begin{eqnarray}\label{WKB_formula}
	\omega^2=V_0 + \Lambda_2(\mathcal{K}^2) + \Lambda_4(\mathcal{K}^2) +....- i\mathcal{K}\sqrt{-2V_2}(1+\Lambda_3(\mathcal{K}^2)+\Lambda_5(\mathcal{K}^2)+....)\,,
\end{eqnarray}
	%%%%%%ppp
where $\mathcal{K}$ takes half-integer value. $\Lambda_k(\mathcal{K}^2)$ is the $k$th order correction term, which depends on the derivative of the effective potential at the position of its maximum. $V_i$ denotes the derivative with respect with $r$ at the position of its maximum. Notice that $V_0$ is the potential itself at the position of its maximum.
	
In order to further improve the accuracy, we consider the WKB formula proposed by Matyjasek and Opala \cite{Matyjasek:2017psv} and use the Pad\'{e} approximant. Then, the WKB formula \eqref{WKB_formula} can be reformulated as
	%%%%%%ppp
\begin{eqnarray}\label{WKB_modify}
	P_k(\epsilon)=V_0 + \Lambda_2(\mathcal{K}^2)\epsilon^2 + \Lambda_4(\mathcal{K}^2) \epsilon^4 +....- i\mathcal{K}\sqrt{-2V_2}(1+\Lambda_3(\mathcal{K}^2)\epsilon^3+\Lambda_5(\mathcal{K}^2)\epsilon^5+....)\,.
	\nonumber
	\\
\end{eqnarray}
	%%%%%%ppp
$P_k(\epsilon)$ called the Pad\'{e} approximant are polynomials of a family of the rational functions
	%%%%%%ppp
\begin{eqnarray}
	P_{\tilde{n}/\tilde{m}}(\epsilon)=\frac{Q_0+Q_1 \epsilon+...+Q_{\tilde{n}}\epsilon^{\tilde{n}}}{R_0+R_1 \epsilon+...+R_{\tilde{m}}\epsilon^{\tilde{m}}}
\end{eqnarray}
	%%%%%%ppp
with $\tilde{n}+\tilde{m}=k$.
The squared frequency is obtained for $\epsilon=1$, i.e., $\omega^2=P_k(1)$.
The improved WKB method with the Pad\'{e} approximant provides a more powerful tool with higher accuracy to find QNMs, especially for $\tilde{n}\approx\tilde{m}\approx k/2$ \cite{Matyjasek:2017psv,Konoplya:2019hlu}.

To estimate the error of the WKB approximation, we define the quantity \cite{Konoplya:2019hlu}
%%%%%%ppp
\begin{eqnarray}\label{error_estimate}
	\triangle_k=\frac{|\omega_{k+1}-\omega_{k-1}|}{2}\,.
\end{eqnarray}
%%%%%%ppp
Then, we can use the following inequation to evaluate the error order \cite{Konoplya:2019hlu}
\begin{eqnarray}
	\triangle_k \gtrsim |\omega-\omega_k |\,,
\end{eqnarray}
%%%%%%ppp
where $\omega$ is the accurate value of the QNM frequency.
	
\begin{figure}[H]
	\center{
		\includegraphics[scale=0.65]{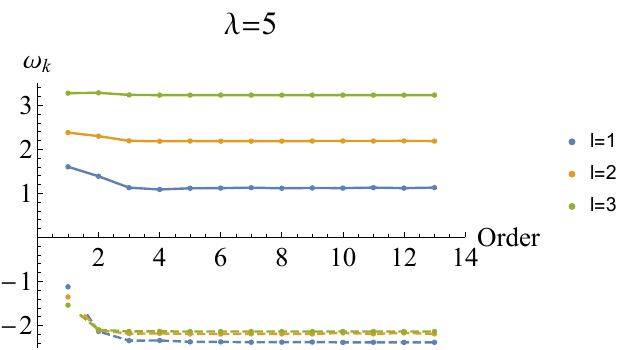}\ \hspace{0.5cm}
		\includegraphics[scale=0.65]{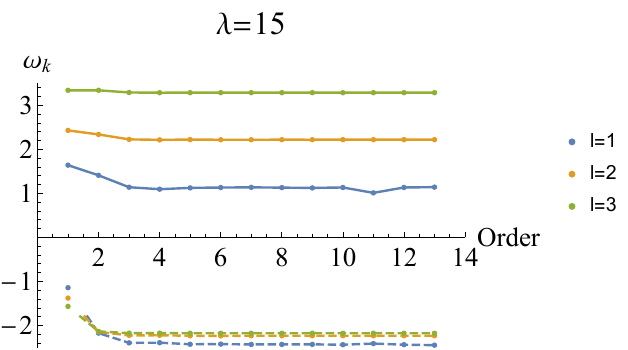}\ \\
		\includegraphics[scale=0.68]{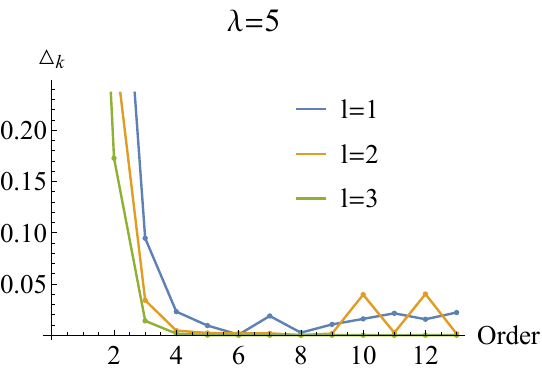}\ \hspace{0.5cm}
		\includegraphics[scale=0.68]{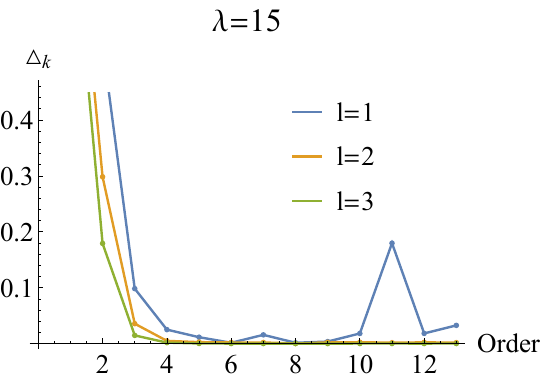}\\
		\caption{The plots above: the real part (solid lines) and imaginary part (dashed line) of the dominant frequency ($n=0$) for different WKB orders.
			The plots below: the error $\triangle_k$ for different WKB orders.}
		\label{error_order}
	      }
\end{figure}

The plots above in Fig.\ref{error_order} show the real and imaginary parts of the dominant frequency ($n=0$) for different WKB orders.
We see that both the real and imaginary parts of the QNM frequency are convergent as the WKB order increases.
Especially, we find that, for our model studied here, it is best to calculate the QNM frequency is the sixth-order WKB approximation, which allows the error estimation of less than $0.1\% $.
	
%%%%%%%%%%
%%%%%%%%%%%	
%%%%%%%%%%
\begin{figure}[H]
	\center{
		\includegraphics[scale=0.7]{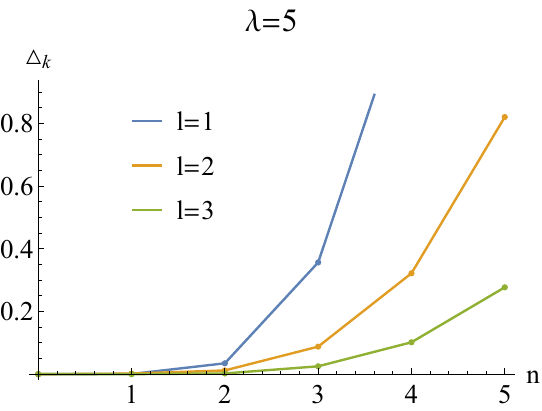}\ \hspace{0.5cm}
		\includegraphics[scale=0.7]{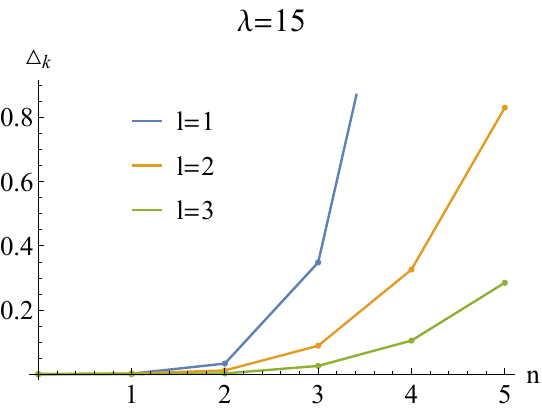}\\
		\caption{The error $\triangle_k$ vs the overtone number $n$ for the sixth-order WKB.}
		\label{error_n}
	}
\end{figure}
%%%%%%%%%%%	

In addition, we also show the error estimation $\triangle_k$ vs the overtone number $n$ for the sixth-order WKB in Fig.\ref{error_n}.
We find that with increasing $n$ for fixed $l$, the error estimation $\triangle_k$ rapidly increases and the WKB method loses its power to find the accurate QNMs. Therefore, the WKB method applies only for $l> n$, which is also pointed out in \cite{Konoplya:2019hlu}.

\section{FILTERING SPURIOUS MODES}\label{filter}

It is well known that the $N\times N$ matrix produces $N$ eigenvalues. However, the majority of the eigenvalues discovered are numerical artifacts, i.e., the spurious modes, with just a handful being correct. Thus, when using the pseudospectral method to find the QNMs, it is crucial to filter out these spurious modes \cite{Jansen:2017oag}. 

%%%%%%%%%%
\begin{figure}[H]
	\center{
		\includegraphics[scale=0.35]{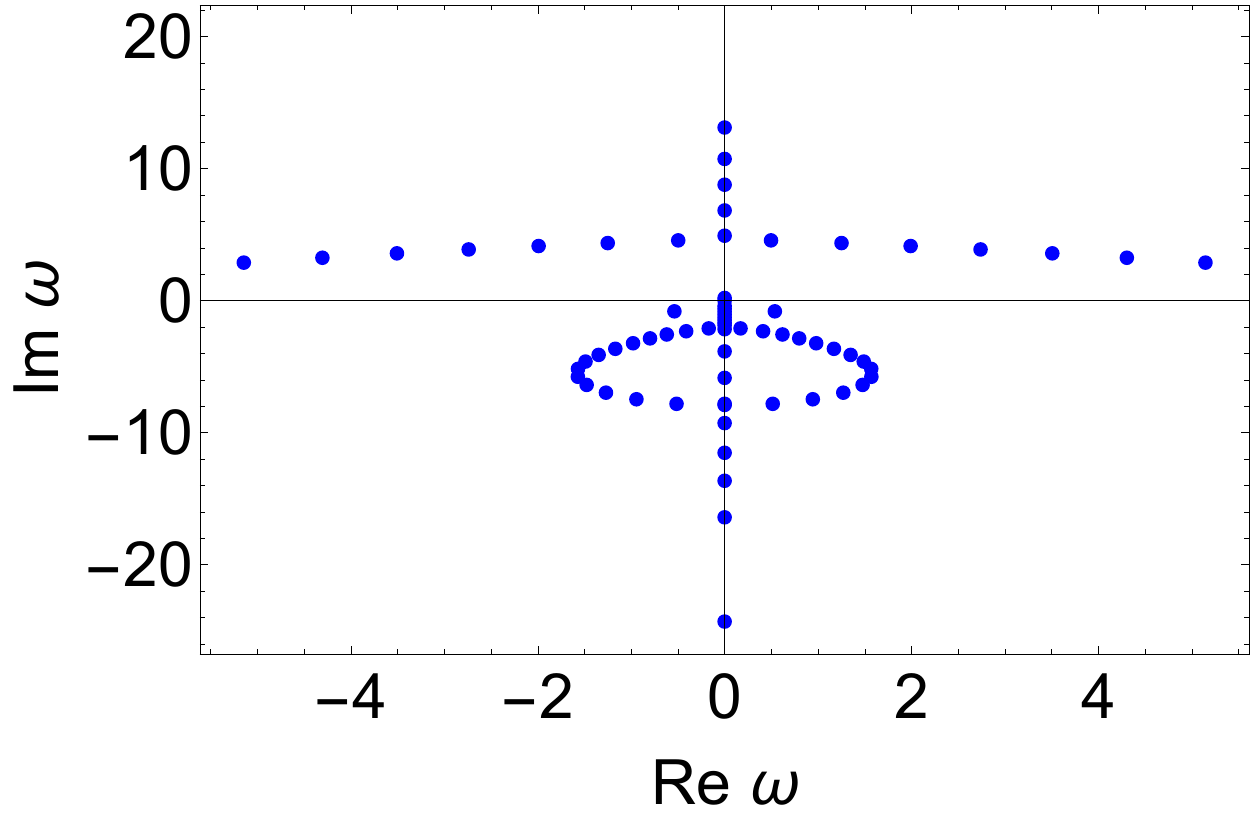}\ \hspace{0.5cm}
		\includegraphics[scale=0.35]{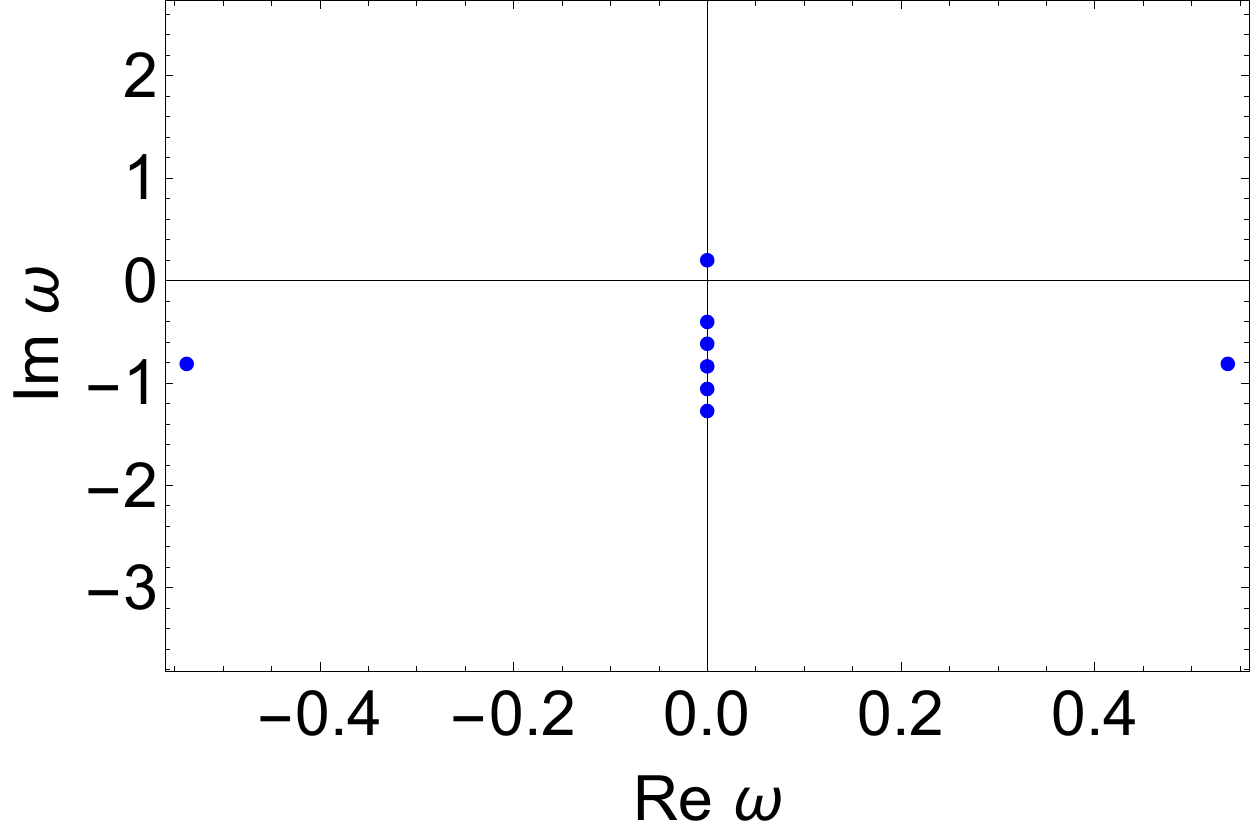}\ \\
		\caption{The QNM spectra of the scalar filed perturbation for $l=0$ and $\lambda=5$. The basis tuple $\{40,40\}$ is used in the left plot, whereas the basis tuples $\{40,40\}$ and $\{100,100\}$ are used in the right plot.}
		\label{eigenvalue_spectral}
	}
\end{figure}
%%%%%%%%%%%	
The first step of filtering\ spurious\ modes is straightforward: repeat the computation at different grid sizes and precisions, then pick the same modes. For convenience, we commonly denote the grid size and precision as $\{N, prec\}$ and refer to them as a basis tuple \cite{Fortuna:2020obg}. The left plot in Fig.\ref{eigenvalue_spectral} displays QNM spectra with a particular grid size and precision $\{40,40\}$. In this $\omega_R-\omega_I$ plan, we see a multitude of modes. Then, we will apply the above-described approach for filtering spurious modes. After performing the computation at two grids sizes and precisions: $\{40,40\}$ and $\{100,100\}$, the majority of modes are found to be rejected. The remaining modes are presented in the right plot of Fig.\ref{eigenvalue_spectral}. On this basis, we have eliminated the majority of spurious modes. However, this is not enough; additional confirmation of whether these modes are genuine is needed. To that purpose, we shall proceed to the second stage: examining the eigenfunction.

\begin{figure}[H]
	\center{
		\includegraphics[scale=0.35]{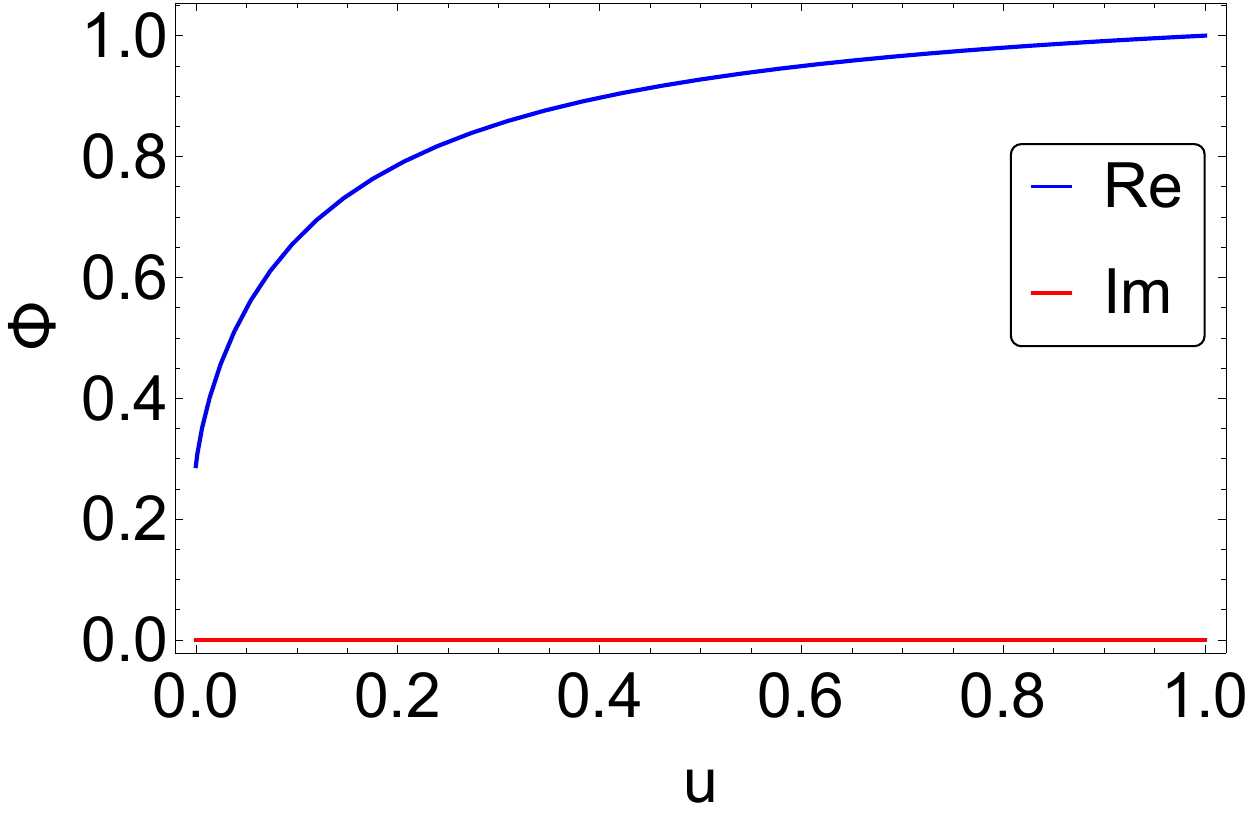}\ \hspace{0.5cm}
		\includegraphics[scale=0.35]{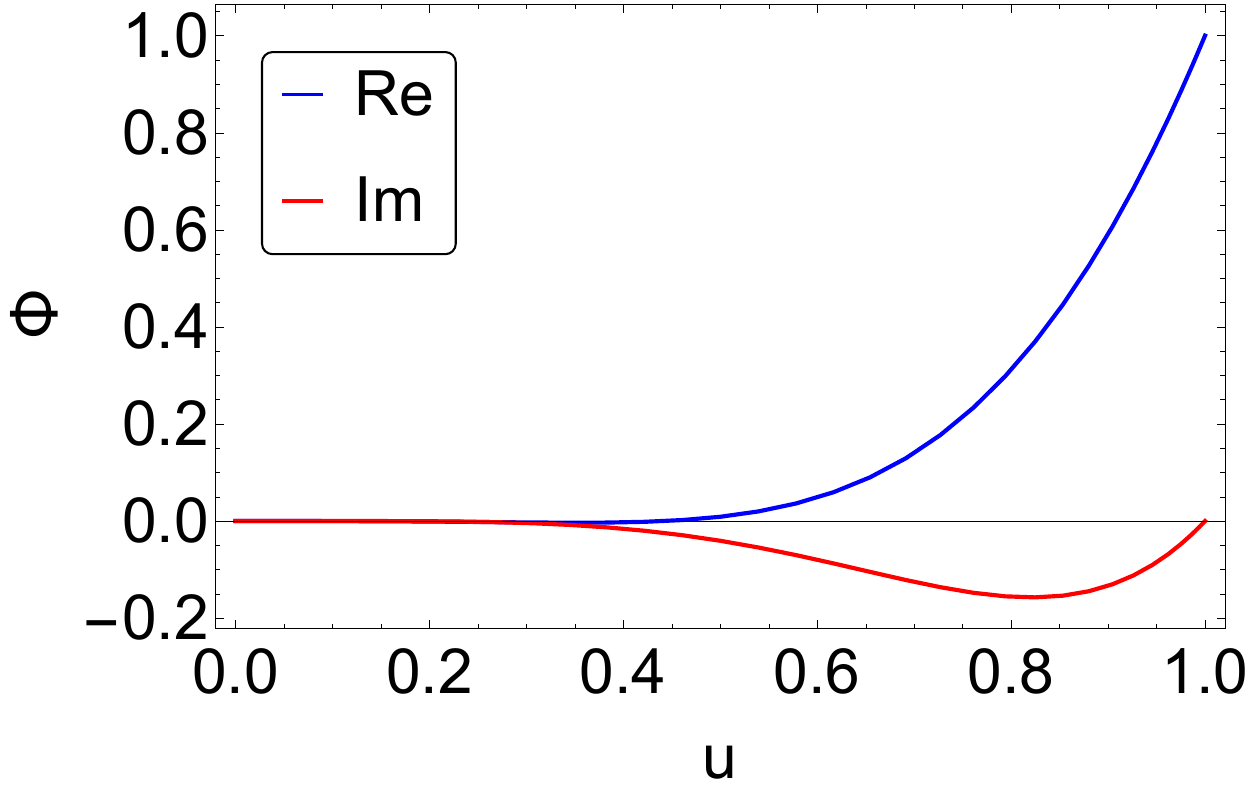}\ \\
		\caption{The eigenfunctions associated with the different modes for $l=0$ and $\lambda=5$. The left plot corresponds to an eigenvalue of $0.19696i$, whereas the right plot corresponds to a value of $0.5374-0.8153i$.}
		\label{eigenfunction}
	}
\end{figure}

The eigenfunction associated with the genuine mode should be smooth, normalized to $1$ at the horizon ($u=1$), and $0$ at the boundary ($u=0$). In Fig.\ref{eigenfunction}, we show the eigenfunctions associated with the modes $0.19696i$ and $0.5374-0.8153i$. Although it survives in any basis tuples, the eigenfunctions associated with the mode $0.19696i$ does not satisfy the boundary conditions (left plot in Fig.\ref{eigenfunction}). As a result, this mode is the spurious mode. The eigenfunction associated with the mode $0.5374-0.8153i$ is shown in the right plot in Fig.\ref{eigenfunction}. This eigenfunction is determined to be smooth and to satisfy the boundary contradictions. Consequently, this is the genuine mode. In addition, we validated the data using the Bernstein spectral method \cite{Fortuna:2020obg}, which is another powerful tool for locating QNMs. All of the QNM results reported in the main body are confirmed using the Bernstein spectral method and verified by the two processes mentioned here to filter out the spurious modes.

%%%%%%%%%%
%%%%%%%%%%%	
	
\bibliographystyle{style1}
\bibliography{Ref}

\providecommand{\href}[2]{#2}\begingroup\raggedright\begin{thebibliography}{10}

\bibitem{LIGOScientific:2016aoc}
{\bf LIGO Scientific, Virgo} Collaboration, B.~P. Abbott et~al., {\it
  {Observation of Gravitational Waves from a Binary Black Hole Merger}},  {\em
  Phys. Rev. Lett.} {\bf 116} (2016), no.~6 061102,
  [\href{http://arxiv.org/abs/1602.03837}{{\tt arXiv:1602.03837}}].

\bibitem{LIGOScientific:2016lio}
{\bf LIGO Scientific, Virgo} Collaboration, B.~P. Abbott et~al., {\it {Tests of
  general relativity with GW150914}},  {\em Phys. Rev. Lett.} {\bf 116} (2016),
  no.~22 221101, [\href{http://arxiv.org/abs/1602.03841}{{\tt
  arXiv:1602.03841}}]. [Erratum: Phys. Rev. Lett. \textbf{121} (2018) 129902 ].

\bibitem{LIGOScientific:2016sjg}
{\bf LIGO Scientific, Virgo} Collaboration, B.~P. Abbott et~al., {\it
  {GW151226: Observation of Gravitational Waves from a 22-Solar-Mass Binary
  Black Hole Coalescence}},  {\em Phys. Rev. Lett.} {\bf 116} (2016), no.~24
  241103, [\href{http://arxiv.org/abs/1606.04855}{{\tt arXiv:1606.04855}}].

\bibitem{EventHorizonTelescope:2019dse}
{\bf Event Horizon Telescope} Collaboration, K.~Akiyama et~al., {\it {First M87
  Event Horizon Telescope Results. I. The Shadow of the Supermassive Black
  Hole}},  {\em Astrophys. J. Lett.} {\bf 875} (2019) L1,
  [\href{http://arxiv.org/abs/1906.11238}{{\tt arXiv:1906.11238}}].

\bibitem{EventHorizonTelescope:2019ths}
{\bf Event Horizon Telescope} Collaboration, K.~Akiyama et~al., {\it {First M87
  Event Horizon Telescope Results. IV. Imaging the Central Supermassive Black
  Hole}},  {\em Astrophys. J. Lett.} {\bf 875} (2019), no.~1 L4,
  [\href{http://arxiv.org/abs/1906.11241}{{\tt arXiv:1906.11241}}].

\bibitem{EventHorizonTelescope:2022xnr}
{\bf Event Horizon Telescope} Collaboration, K.~Akiyama et~al., {\it {First
  Sagittarius A* Event Horizon Telescope Results. I. The Shadow of the
  Supermassive Black Hole in the Center of the Milky Way}},  {\em Astrophys. J.
  Lett.} {\bf 930} (2022), no.~2 L12.

\bibitem{EventHorizonTelescope:2022xqj}
{\bf Event Horizon Telescope} Collaboration, K.~Akiyama et~al., {\it {First
  Sagittarius A* Event Horizon Telescope Results. VI. Testing the Black Hole
  Metric}},  {\em Astrophys. J. Lett.} {\bf 930} (2022), no.~2 L17.

\bibitem{Weyl:1918pdp}
H.~Weyl, {\it {Reine In nitesimalgeometrie}},  {\em Math. Z.} {\bf 2} (1918),
  no.~3-4 384--411.

\bibitem{PhysRevD.16.953}
K.~S. Stelle, {\it Renormalization of higher-derivative quantum gravity},  {\em
  Phys. Rev. D} {\bf 16} (Aug, 1977) 953--969.

\bibitem{Faria:2015vea}
F.~F. Faria, {\it {Quantum massive conformal gravity}},  {\em Eur. Phys. J. C}
  {\bf 76} (2016), no.~4 188, [\href{http://arxiv.org/abs/1503.04355}{{\tt
  arXiv:1503.04355}}].

\bibitem{BERGSHOEFF1981173}
{\it Extended conformal supergravity},  {\em Nucl. Phys. B} {\bf 182} (1981),
  no.~1 173--204.

\bibitem{deWit:1980lyi}
B.~de~Wit, J.~W. van Holten, and A.~Van~Proeyen, {\it {Structure of N=2
  Supergravity}},  {\em Nucl. Phys. B} {\bf 184} (1981) 77. [Erratum: Nucl.
  Phys. B \textbf{222} (1983) 516 ].

\bibitem{Maldacena:2011mk}
J.~Maldacena, {\it {Einstein Gravity from Conformal Gravity}},
  \href{http://arxiv.org/abs/1105.5632}{{\tt arXiv:1105.5632}}.

\bibitem{Anastasiou:2016jix}
G.~Anastasiou and R.~Olea, {\it {From conformal to Einstein Gravity}},  {\em
  Phys. Rev. D} {\bf 94} (2016), no.~8 086008,
  [\href{http://arxiv.org/abs/1608.07826}{{\tt arXiv:1608.07826}}].

\bibitem{Konoplya:2020fwg}
R.~A. Konoplya, {\it {Conformal Weyl gravity via two stages of quasinormal
  ringing and late-time behavior}},  {\em Phys. Rev. D} {\bf 103} (2021), no.~4
  044033, [\href{http://arxiv.org/abs/2012.13020}{{\tt arXiv:2012.13020}}].

\bibitem{Mannheim:1988dj}
P.~D. Mannheim and D.~Kazanas, {\it {Exact Vacuum Solution to Conformal Weyl
  Gravity and Galactic Rotation Curves}},  {\em Astrophys. J.} {\bf 342} (1989)
  635--638.

\bibitem{Mannheim:2005bfa}
P.~D. Mannheim, {\it {Alternatives to dark matter and dark energy}},  {\em
  Prog. Part. Nucl. Phys.} {\bf 56} (2006) 340--445,
  [\href{http://arxiv.org/abs/astro-ph/0505266}{{\tt astro-ph/0505266}}].

\bibitem{Robert:2013Entrp}
R.~{Nesbet}, {\it {Conformal Gravity: Dark Matter and Dark Energy}},  {\em
  Entropy} {\bf 15} (Jan., 2013) 162--176,
  [\href{http://arxiv.org/abs/1208.4972}{{\tt arXiv:1208.4972}}].

\bibitem{Mannheim:1990ya}
P.~D. Mannheim and D.~Kazanas, {\it {Solutions to the Kerr and Kerr-Newman
  problems in fourth order conformal Weyl gravity}},  {\em Phys. Rev. D} {\bf
  44} (1991) 417--423.

\bibitem{Tanhayi:2011dh}
M.~R. Tanhayi, M.~Fathi, and M.~V. Takook, {\it {Observable Quantities in Weyl
  Gravity}},  {\em Mod. Phys. Lett. A} {\bf 26} (2011) 2403--2410,
  [\href{http://arxiv.org/abs/1108.6157}{{\tt arXiv:1108.6157}}].

\bibitem{Payandeh:2012mj}
F.~Payandeh and M.~Fathi, {\it {Spherical Solutions due to the Exterior
  Geometry of a Charged Weyl Black Hole}},  {\em Int. J. Theor. Phys.} {\bf 51}
  (2012) 2227--2236, [\href{http://arxiv.org/abs/1202.2415}{{\tt
  arXiv:1202.2415}}].

\bibitem{Berti:2009kk}
E.~Berti, V.~Cardoso, and A.~O. Starinets, {\it {Quasinormal modes of black
  holes and black branes}},  {\em Class. Quant. Grav.} {\bf 26} (2009) 163001,
  [\href{http://arxiv.org/abs/0905.2975}{{\tt arXiv:0905.2975}}].

\bibitem{Berti:2005ys}
E.~Berti, V.~Cardoso, and C.~M. Will, {\it {On gravitational-wave spectroscopy
  of massive black holes with the space interferometer LISA}},  {\em Phys. Rev.
  D} {\bf 73} (2006) 064030, [\href{http://arxiv.org/abs/gr-qc/0512160}{{\tt
  gr-qc/0512160}}].

\bibitem{Berti:2018vdi}
E.~Berti, K.~Yagi, H.~Yang, and N.~Yunes, {\it {Extreme Gravity Tests with
  Gravitational Waves from Compact Binary Coalescences: (II) Ringdown}},  {\em
  Gen. Rel. Grav.} {\bf 50} (2018), no.~5 49,
  [\href{http://arxiv.org/abs/1801.03587}{{\tt arXiv:1801.03587}}].

\bibitem{Hawking:1975}
S.~W. Hawking, {\it {Particle Creation by Black Holes}},  {\em Commun. Math.
  Phys.} {\bf 43} (1975) 199--220.

\bibitem{PhysRevD.13.198}
D.~N. Page, {\it Particle emission rates from a black hole: Massless particles
  from an uncharged, nonrotating hole},  {\em Phys. Rev. D} {\bf 13} (Jan,
  1976) 198--206.

\bibitem{Harris:2003eg}
C.~M. Harris and P.~Kanti, {\it {Hawking radiation from a (4+n)-dimensional
  black hole: Exact results for the Schwarzschild phase}},  {\em JHEP} {\bf 10}
  (2003) 014, [\href{http://arxiv.org/abs/hep-ph/0309054}{{\tt
  hep-ph/0309054}}].

\bibitem{Zhang:2020qam}
C.-Y. Zhang, P.-C. Li, and M.~Guo, {\it {Greybody factor and power spectra of
  the Hawking radiation in the $4D$
  Einstein\textendash{}Gauss\textendash{}Bonnet de-Sitter gravity}},  {\em Eur.
  Phys. J. C} {\bf 80} (2020), no.~9 874,
  [\href{http://arxiv.org/abs/2003.13068}{{\tt arXiv:2003.13068}}].

\bibitem{Konoplya:2019hml}
R.~A. Konoplya, A.~F. Zinhailo, and Z.~Stuchl\'\i{}k, {\it {Quasinormal modes,
  scattering, and Hawking radiation in the vicinity of an
  Einstein-dilaton-Gauss-Bonnet black hole}},  {\em Phys. Rev. D} {\bf 99}
  (2019), no.~12 124042, [\href{http://arxiv.org/abs/1903.03483}{{\tt
  arXiv:1903.03483}}].

\bibitem{Konoplya:2019ppy}
R.~A. Konoplya and A.~F. Zinhailo, {\it {Hawking radiation of non-Schwarzschild
  black holes in higher derivative gravity: a crucial role of grey-body
  factors}},  {\em Phys. Rev. D} {\bf 99} (2019), no.~10 104060,
  [\href{http://arxiv.org/abs/1904.05341}{{\tt arXiv:1904.05341}}].

\bibitem{Konoplya:2020jgt}
R.~A. Konoplya, A.~F. Zinhailo, and Z.~Stuchlik, {\it {Quasinormal modes and
  Hawking radiation of black holes in cubic gravity}},  {\em Phys. Rev. D} {\bf
  102} (2020), no.~4 044023, [\href{http://arxiv.org/abs/2006.10462}{{\tt
  arXiv:2006.10462}}].

\bibitem{Konoplya:2020cbv}
R.~A. Konoplya and A.~F. Zinhailo, {\it {Grey-body factors and Hawking
  radiation of black holes in $4D$ Einstein-Gauss-Bonnet gravity}},  {\em Phys.
  Lett. B} {\bf 810} (2020) 135793,
  [\href{http://arxiv.org/abs/2004.02248}{{\tt arXiv:2004.02248}}].

\bibitem{Konoplya:2021ube}
R.~A. Konoplya, {\it {Black holes in galactic centers: Quasinormal ringing,
  grey-body factors and Unruh temperature}},  {\em Phys. Lett. B} {\bf 823}
  (2021) 136734, [\href{http://arxiv.org/abs/2109.01640}{{\tt
  arXiv:2109.01640}}].

\bibitem{Guo:2020blq}
H.~Guo, H.~Liu, X.-M. Kuang, and B.~Wang, {\it {Acoustic black hole in
  Schwarzschild spacetime: quasi-normal modes, analogous Hawking radiation and
  shadows}},  {\em Phys. Rev. D} {\bf 102} (2020) 124019,
  [\href{http://arxiv.org/abs/2007.04197}{{\tt arXiv:2007.04197}}].

\bibitem{Ling:2021vgk}
R.~Ling, H.~Guo, H.~Liu, X.-M. Kuang, and B.~Wang, {\it {Shadow and
  near-horizon characteristics of the acoustic charged black hole in curved
  spacetime}},  {\em Phys. Rev. D} {\bf 104} (2021), no.~10 104003,
  [\href{http://arxiv.org/abs/2107.05171}{{\tt arXiv:2107.05171}}].

\bibitem{Syu:2022cws}
W.-C. Syu, D.-S. Lee, and C.-Y. Lin, {\it {Analogous Hawking radiation and
  quantum entanglement in two-component Bose-Einstein condensates: the gapped
  excitations}},  \href{http://arxiv.org/abs/2204.10790}{{\tt
  arXiv:2204.10790}}.

\bibitem{Fathi:2020sey}
M.~Fathi, M.~Kariminezhad, M.~Olivares, and J.~R. Villanueva, {\it {Motion of
  massive particles around a charged Weyl black hole and the geodetic
  precession of orbiting gyroscopes}},  {\em Eur. Phys. J. C} {\bf 80} (2020),
  no.~5 377, [\href{http://arxiv.org/abs/2009.03399}{{\tt arXiv:2009.03399}}].

\bibitem{Fathi:2019jgd}
M.~Fathi, M.~Olivares, and J.~R. Villanueva, {\it {Classical tests on a charged
  Weyl black hole: bending of light, Shapiro delay and Sagnac effect}},  {\em
  Eur. Phys. J. C} {\bf 80} (2020), no.~1 51,
  [\href{http://arxiv.org/abs/1910.12811}{{\tt arXiv:1910.12811}}].

\bibitem{Fathi:2020otm}
M.~Fathi and J.~R. Villanueva, {\it {The role of elliptic integrals in
  calculating the gravitational lensing of a charged Weyl black hole surrounded
  by plasma}},  \href{http://arxiv.org/abs/2009.03402}{{\tt arXiv:2009.03402}}.

\bibitem{Fathi:2020sfw}
M.~Fathi, M.~Olivares, and J.~R. Villanueva, {\it {Gravitational Rutherford
  scattering of electrically charged particles from a charged Weyl black
  hole}},  {\em Eur. Phys. J. Plus} {\bf 136} (2021), no.~4 420,
  [\href{http://arxiv.org/abs/2009.03404}{{\tt arXiv:2009.03404}}].

\bibitem{Chandrasekhar}
S.~Chandrasekhar, {\it {The mathematical theory of black holes. Oxford classic
  texts in the physical sciences}},  {\em Oxford Univ. Press, Oxford} (2002).

\bibitem{Nollert:1999ji}
H.-P. Nollert, {\it {TOPICAL REVIEW: Quasinormal modes: the characteristic
  `sound' of black holes and neutron stars}},  {\em Class. Quant. Grav.} {\bf
  16} (1999) R159--R216.

\bibitem{Konoplya:2011qq}
R.~A. Konoplya and A.~Zhidenko, {\it {Quasinormal modes of black holes: From
  astrophysics to string theory}},  {\em Rev. Mod. Phys.} {\bf 83} (2011)
  793--836, [\href{http://arxiv.org/abs/1102.4014}{{\tt arXiv:1102.4014}}].

\bibitem{Kokkotas:1999bd}
K.~D. Kokkotas and B.~G. Schmidt, {\it {Quasinormal modes of stars and black
  holes}},  {\em Living Rev. Rel.} {\bf 2} (1999) 2,
  [\href{http://arxiv.org/abs/gr-qc/9909058}{{\tt gr-qc/9909058}}].

\bibitem{Boyd:Chebyshev}
J.~P. Boyd, {\it {Chebyshev $\&$ Fourier Spectral Methods}},  {\em Courier
  Dover Publications}.

\bibitem{Jansen:2017oag}
A.~Jansen, {\it {Overdamped modes in Schwarzschild-de Sitter and a Mathematica
  package for the numerical computation of quasinormal modes}},  {\em Eur.
  Phys. J. Plus} {\bf 132} (2017), no.~12 546,
  [\href{http://arxiv.org/abs/1709.09178}{{\tt arXiv:1709.09178}}].

\bibitem{Wu:2018vlj}
J.-P. Wu and P.~Liu, {\it {Quasi-normal modes of holographic system with Weyl
  correction and momentum dissipation}},  {\em Phys. Lett. B} {\bf 780} (2018)
  616--621, [\href{http://arxiv.org/abs/1804.10897}{{\tt arXiv:1804.10897}}].

\bibitem{Fu:2018yqx}
G.~Fu and J.-P. Wu, {\it {EM Duality and Quasinormal Modes from Higher
  Derivatives with Homogeneous Disorder}},  {\em Adv. High Energy Phys.} {\bf
  2019} (2019) 5472310, [\href{http://arxiv.org/abs/1812.11522}{{\tt
  arXiv:1812.11522}}].

\bibitem{Xiong:2021cth}
W.~Xiong, P.~Liu, C.-Y. Zhang, and C.~Niu, {\it {Quasi-normal modes of the
  Einstein-Maxwell-aether Black Hole}},
  \href{http://arxiv.org/abs/2112.12523}{{\tt arXiv:2112.12523}}.

\bibitem{Liu:2021fzr}
P.~Liu, C.~Niu, and C.-Y. Zhang, {\it {Linear instability of charged massless
  scalar perturbation in regularized 4D charged Einstein-Gauss-Bonnet anti
  de-Sitter black holes}},  {\em Chin. Phys. C} {\bf 45} (2021), no.~2 025111.

\bibitem{Liu:2021zmi}
P.~Liu, C.~Niu, and C.-Y. Zhang, {\it {Instability of regularized 4D charged
  Einstein-Gauss-Bonnet de-Sitter black holes}},  {\em Chin. Phys. C} {\bf 45}
  (2021), no.~2 025104.

\bibitem{Jaramillo:2020tuu}
J.~L. Jaramillo, R.~Panosso~Macedo, and L.~Al~Sheikh, {\it {Pseudospectrum and
  Black Hole Quasinormal Mode Instability}},  {\em Phys. Rev. X} {\bf 11}
  (2021), no.~3 031003, [\href{http://arxiv.org/abs/2004.06434}{{\tt
  arXiv:2004.06434}}].

\bibitem{Jaramillo:2021tmt}
J.~L. Jaramillo, R.~Panosso~Macedo, and L.~A. Sheikh, {\it {Gravitational wave
  signatures of black hole quasi-normal mode instability}},
  \href{http://arxiv.org/abs/2105.03451}{{\tt arXiv:2105.03451}}.

\bibitem{Destounis:2021lum}
K.~Destounis, R.~P. Macedo, E.~Berti, V.~Cardoso, and J.~L. Jaramillo, {\it
  {Pseudospectrum of Reissner-Nordstr\"om black holes: Quasinormal mode
  instability and universality}},  {\em Phys. Rev. D} {\bf 104} (2021), no.~8
  084091, [\href{http://arxiv.org/abs/2107.09673}{{\tt arXiv:2107.09673}}].

\bibitem{Cardoso:2017soq}
V.~Cardoso, J.~a.~L. Costa, K.~Destounis, P.~Hintz, and A.~Jansen, {\it
  {Quasinormal modes and Strong Cosmic Censorship}},  {\em Phys. Rev. Lett.}
  {\bf 120} (2018), no.~3 031103, [\href{http://arxiv.org/abs/1711.10502}{{\tt
  arXiv:1711.10502}}].

\bibitem{Konoplya:2022gjp}
R.~A. Konoplya, {\it {Further clarification on quasinormal modes/circular null
  geodesics correspondence}},  \href{http://arxiv.org/abs/2210.08373}{{\tt
  arXiv:2210.08373}}.

\bibitem{Zhidenko:2003wq}
A.~Zhidenko, {\it {Quasinormal modes of Schwarzschild de Sitter black holes}},
  {\em Class. Quant. Grav.} {\bf 21} (2004) 273--280,
  [\href{http://arxiv.org/abs/gr-qc/0307012}{{\tt gr-qc/0307012}}].

\bibitem{Konoplya:2003ii}
R.~A. Konoplya, {\it {Quasinormal behavior of the d-dimensional Schwarzschild
  black hole and higher order WKB approach}},  {\em Phys. Rev. D} {\bf 68}
  (2003) 024018, [\href{http://arxiv.org/abs/gr-qc/0303052}{{\tt
  gr-qc/0303052}}].

\bibitem{Konoplya:2019hlu}
R.~A. Konoplya, A.~Zhidenko, and A.~F. Zinhailo, {\it {Higher order WKB formula
  for quasinormal modes and grey-body factors: recipes for quick and accurate
  calculations}},  {\em Class. Quant. Grav.} {\bf 36} (2019) 155002,
  [\href{http://arxiv.org/abs/1904.10333}{{\tt arXiv:1904.10333}}].

\bibitem{Abdalla:2010nq}
E.~Abdalla, C.~E. Pellicer, J.~de~Oliveira, and A.~B. Pavan, {\it Phase
  transitions and regions of stability in reissner-nordstr\"om holographic
  superconductors},  {\em Phys. Rev. D} {\bf 82} (2010) 124033,
  [\href{http://arxiv.org/abs/1010.2806}{{\tt arXiv:1010.2806}}].

\bibitem{Zhu:2014sya}
Z.~Zhu, S.-J. Zhang, C.~E. Pellicer, B.~Wang, and E.~Abdalla, {\it Stability of
  reissner-nordstr\"om black hole in de sitter background under charged scalar
  perturbation},  {\em Phys. Rev. D} {\bf 90} (2014), no.~4 044042,
  [\href{http://arxiv.org/abs/1405.4931}{{\tt arXiv:1405.4931}}]. [Addendum:
  Phys. Rev. D 90 (2014) 049904 ].

\bibitem{Lin:2022owb}
K.~Lin and W.-L. Qian, {\it {Echoes in star quasinormal modes using an
  alternative finite difference method}},
  \href{http://arxiv.org/abs/2204.09531}{{\tt arXiv:2204.09531}}.

\bibitem{Brady:1996za}
P.~R. Brady, C.~M. Chambers, W.~Krivan, and P.~Laguna, {\it {Telling tails in
  the presence of a cosmological constant}},  {\em Phys. Rev. D} {\bf 55}
  (1997) 7538--7545, [\href{http://arxiv.org/abs/gr-qc/9611056}{{\tt
  gr-qc/9611056}}].

\bibitem{Brady:1999wd}
P.~R. Brady, C.~M. Chambers, W.~G. Laarakkers, and E.~Poisson, {\it {Radiative
  falloff in Schwarzschild-de Sitter space-time}},  {\em Phys. Rev. D} {\bf 60}
  (1999) 064003, [\href{http://arxiv.org/abs/gr-qc/9902010}{{\tt
  gr-qc/9902010}}].

\bibitem{Molina:2003dc}
C.~Molina, D.~Giugno, E.~Abdalla, and A.~Saa, {\it {Field propagation in de
  Sitter black holes}},  {\em Phys. Rev. D} {\bf 69} (2004) 104013,
  [\href{http://arxiv.org/abs/gr-qc/0309079}{{\tt gr-qc/0309079}}].

\bibitem{PhysRevD.35.3621}
S.~Iyer and C.~M. Will, {\it Black-hole normal modes: A wkb approach. i.
  foundations and application of a higher-order wkb analysis of
  potential-barrier scattering},  {\em Phys. Rev. D} {\bf 35} (Jun, 1987)
  3621--3631.

\bibitem{1985ApJ291L33S}
B.~Schutz and C.~Will, {\it {Black hole normal modes - A semianalytic approach,
  Astrophys. J. Lett }},  {\em Astrophys. J. Lett} {\bf 291} (Apr., 1985)
  L33--L36.

\bibitem{Kanti:2004nr}
P.~Kanti, {\it {Black holes in theories with large extra dimensions: A
  Review}},  {\em Int. J. Mod. Phys. A} {\bf 19} (2004) 4899--4951,
  [\href{http://arxiv.org/abs/hep-ph/0402168}{{\tt hep-ph/0402168}}].

\bibitem{Hawking:1975vcx}
S.~W. Hawking, {\it {Particle Creation by Black Holes}},  {\em Commun. Math.
  Phys.} {\bf 43} (1975) 199--220. [Erratum: Commun. Math. Phys. \textbf{46}
  (1976) 206 ].

\bibitem{Lagos:2020oek}
M.~Lagos, P.~G. Ferreira, and O.~J. Tattersall, {\it {Anomalous decay rate of
  quasinormal modes}},  {\em Phys. Rev. D} {\bf 101} (2020), no.~8 084018,
  [\href{http://arxiv.org/abs/2002.01897}{{\tt arXiv:2002.01897}}].

\bibitem{Aragon:2020teq}
A.~Arag\'on, R.~B\'ecar, P.~A. Gonz\'alez, and Y.~V\'asquez, {\it {Massive
  Dirac quasinormal modes in Schwarzschild\textendash{}de Sitter black holes:
  Anomalous decay rate and fine structure}},  {\em Phys. Rev. D} {\bf 103}
  (2021), no.~6 064006, [\href{http://arxiv.org/abs/2009.09436}{{\tt
  arXiv:2009.09436}}].

\bibitem{Fontana:2020syy}
R.~D.~B. Fontana, P.~A. Gonz\'alez, E.~Papantonopoulos, and Y.~V\'asquez, {\it
  {Anomalous decay rate of quasinormal modes in Reissner-Nordstr\"om black
  holes}},  {\em Phys. Rev. D} {\bf 103} (2021), no.~6 064005,
  [\href{http://arxiv.org/abs/2011.10620}{{\tt arXiv:2011.10620}}].

\bibitem{Doneva:2017bvd}
D.~D. Doneva and S.~S. Yazadjiev, {\it {New Gauss-Bonnet Black Holes with
  Curvature-Induced Scalarization in Extended Scalar-Tensor Theories}},  {\em
  Phys. Rev. Lett.} {\bf 120} (2018), no.~13 131103,
  [\href{http://arxiv.org/abs/1711.01187}{{\tt arXiv:1711.01187}}].

\bibitem{Silva:2017uqg}
H.~O. Silva, J.~Sakstein, L.~Gualtieri, T.~P. Sotiriou, and E.~Berti, {\it
  {Spontaneous scalarization of black holes and compact stars from a
  Gauss-Bonnet coupling}},  {\em Phys. Rev. Lett.} {\bf 120} (2018), no.~13
  131104, [\href{http://arxiv.org/abs/1711.02080}{{\tt arXiv:1711.02080}}].

\bibitem{Herdeiro:2018wub}
C.~A.~R. Herdeiro, E.~Radu, N.~Sanchis-Gual, and J.~A. Font, {\it {Spontaneous
  Scalarization of Charged Black Holes}},  {\em Phys. Rev. Lett.} {\bf 121}
  (2018), no.~10 101102, [\href{http://arxiv.org/abs/1806.05190}{{\tt
  arXiv:1806.05190}}].

\bibitem{Yang:2021yoe}
Z.-H. Yang, G.~Fu, X.-M. Kuang, and J.-P. Wu, {\it {Instability of de-Sitter
  black hole with massive scalar field coupled to Gauss\textendash{}Bonnet
  invariant and the scalarized black holes}},  {\em Eur. Phys. J. C} {\bf 82}
  (2022), no.~10 868, [\href{http://arxiv.org/abs/2112.15052}{{\tt
  arXiv:2112.15052}}].

\bibitem{Iyer:1986np}
S.~Iyer and C.~M. Will, {\it {Black Hole Normal Modes: A {WKB} Approach. 1.
  Foundations and Application of a Higher Order {WKB} Analysis of Potential
  Barrier Scattering}},  {\em Phys. Rev. D} {\bf 35} (1987) 3621.

\bibitem{Guinn:1989bn}
J.~W. Guinn, C.~M. Will, Y.~Kojima, and B.~F. Schutz, {\it {High Overtone
  Normal Modes of Schwarzschild Black Holes}},  {\em Class. Quant. Grav.} {\bf
  7} (1990) L47.

\bibitem{Konoplya:2004ip}
R.~A. Konoplya, {\it {Quasinormal modes of the Schwarzschild black hole and
  higher order WKB approach}},  {\em J. Phys. Stud.} {\bf 8} (2004) 93--100.

\bibitem{Matyjasek:2017psv}
J.~Matyjasek and M.~Opala, {\it {Quasinormal modes of black holes. The improved
  semianalytic approach}},  {\em Phys. Rev. D} {\bf 96} (2017), no.~2 024011,
  [\href{http://arxiv.org/abs/1704.00361}{{\tt arXiv:1704.00361}}].

\bibitem{Fortuna:2020obg}
S.~Fortuna and I.~Vega, {\it {Bernstein spectral method for quasinormal modes
  and other eigenvalue problems}},  \href{http://arxiv.org/abs/2003.06232}{{\tt
  arXiv:2003.06232}}.

\end{thebibliography}\endgroup
\end{document}